%% file: main.tex
\documentclass{vgtc}                          

\ifpdf
  \pdfoutput=1\relax                   
  \usepackage{graphicx}                
  \DeclareGraphicsExtensions{.pdf,.png,.jpg,.jpeg} 
\else
  \usepackage{graphicx}                
  \DeclareGraphicsExtensions{.eps}     
\fi%

\graphicspath{{figures/}{pictures/}{images/}{./}} 

\usepackage{microtype}                 
\PassOptionsToPackage{warn}{textcomp}  
\usepackage{textcomp}                  
\usepackage{mathptmx}                  
\usepackage{times}                     
\usepackage{cite}                      
\usepackage{tabu}                      
\usepackage{booktabs}                  
\usepackage{float}
\usepackage{adjustbox}
\usepackage{xr}
\usepackage{hyperref}
\usepackage[table,xcdraw]{xcolor}
\usepackage[normalem]{ulem}

 \newcommand{\kh}[1]{}
 \newcommand{\amanat}[1]{}
 \newcommand{\sid}[1]{}
 \newcommand{\remove}[1]{} 
 \newcommand{\update}[1]{\textcolor{black}{#1}}

\newcommand{\degree}{\textbf{$^{\circ}$}}
\usepackage{academicons}

\onlineid{0}

\vgtccategory{Research}

\vgtcinsertpkg

\title{Exploring Bi-Manual Teleportation in Virtual Reality}

\author{Siddhanth Raja Sindhupathiraja \thanks{e-mail: cs5200443@iitd.ac.in}\\ %
        \scriptsize Department of Computer Science and Engineering
        \\\scriptsize Indian Institute of Technology Delhi
        \\\scriptsize New Delhi, Delhi, India %
\and A K M Amanat Ullah  \thanks{e-mail: amanat7@student.ubc.ca}\\ %
     \scriptsize University of British Columbia \\\scriptsize Okanagan, BC, Canada %
\and William Delamare \thanks{e-mail: w.delamare@estia.fr}\\ %
     \scriptsize Univ. Bordeaux, ESTIA-Institute of Technology, EstiaR \\\scriptsize F-64210 Bidart, France%
\and Khalad Hasan  \thanks{e-mail: khalad.hasan@ubc.ca}\\ %
      \scriptsize University of British Columbia \\\scriptsize Okanagan, BC, Canada}

\abstract{
Teleportation, a widely-used locomotion technique in Virtual Reality (VR), allows instantaneous movement within VR environments. Enhanced hand tracking in modern VR headsets has popularized hands-only teleportation methods, which eliminate the need for physical controllers. However, these techniques have not fully explored the potential of bi-manual input, where each hand plays a distinct role in teleportation: one controls the teleportation point and the other confirms selections. Additionally, the influence of users' posture, whether sitting or standing, on these techniques remains unexplored. Furthermore, previous teleportation evaluations lacked assessments based on established human motor models such as Fitts' Law. To address these gaps, we conducted a user study (N=20) to evaluate bi-manual pointing performance in VR teleportation tasks, considering both sitting and standing postures. We proposed a variation of the Fitts' Law model to accurately assess users' teleportation performance. We designed and evaluated various bi-manual teleportation techniques, comparing them to uni-manual and dwell-based techniques. Results showed that bi-manual techniques, particularly when the dominant hand is used for pointing and the non-dominant hand for selection, enable faster teleportation compared to other methods. Furthermore, bi-manual and dwell techniques proved significantly more accurate than uni-manual teleportation. Moreover, our proposed Fitts' Law variation more accurately predicted users' teleportation performance compared to existing models. Finally, we developed a set of guidelines for designers to enhance VR teleportation experiences and optimize user interactions. 
}

\CCScatlist{
  \CCScatTwelve{Human-centered computing}{Human computer interaction}{Interaction paradigms}{Virtual reality};
  \CCScatTwelve{Human-centered computing}{Human computer interaction}{Interaction techniques}{Gestural input}
}

\begin{document}


\maketitle

\input{1_introduction}

\input{2_background}

\input{3_design-space}

\input{5_exp-protocol}

\input{6_exp-results}

\input{7_guidelines}
\input{8_conclusion}



\bibliographystyle{abbrv-doi-hyperref}
\bibliography{main}


\end{document}

%% file: 1_introduction.tex
\section{Introduction}
Teleportation, the technique enabling instant movement within the VR environment, has become an industry standard, appreciated for its adaptability in confined spaces \cite{cherni2020literature, folmer2021teleportation} and its \remove{reduced} \update{lower} motion sickness impact compared to other methods \cite{clifton2020effects, christou2017steering}. In the past, teleportation involved using controllers, where users controlled a teleportation parabola by moving the controller and teleported to a destination by pressing a hardware button \cite{bhandari2018teleportation, funk2019assessing, langbehn2018evaluation}. However, with advanced hand-tracking in VR headsets, hands-only teleportation has emerged as a viable alternative to traditional controllers \cite{schafer2021controlling, chowdhury2022wriarm}. This method allows users to intuitively navigate virtual environments by controlling the pointer with hand movement and \remove{triggering}\update{confirming} selections using hand gestures (e.g., closing the index finger). Given the prevalence of teleportation, evaluating ergonomic factors such as hand input (uni-manual and bi-manual) and user posture (sitting or standing) is essential. \update{Furthermore, VR researchers use Fitts' law as a guiding framework to evaluate their interface designs or interaction techniques \cite{triantafyllidis2021challenges}.
As teleportation techniques require pointing at targets, applying Fitts' law is critical to predict users' teleportation performance in VR.}

Researchers explored diverse hands-only teleportation techniques that leverage hand movements for navigation within virtual environments \cite{huang2019design, khundam2015first, schafer2021controlling, zhang2017double}. These approaches demonstrated advantages over controller-based methods, including reduced motion sickness and workloads, while ensuring an intuitive and immersive experience \cite{khundam2015first, pai2017armswing}. Indeed, the effectiveness of hand-based teleportation is influenced by individual factors, such as hand dominance and user posture (e.g., sitting vs. standing). For instance, users may prefer to use their dominant hand over non-dominant hand for pointing at targets. 
However, \remove{to our best knowledge}\update{to the best of our knowledge}, the only research that explored uni-manual and bi-manual teleportation was by \remove{Schafer} \update{Sch{\"a}fer} et al. \cite{schafer2021controlling}, where they performed a limited comparison of teleportation techniques considering the right hand for pointing, neglecting the left hand. Furthermore, their work primarily focused on fundamental tasks involving large targets placed at constant distances on the fixed elevation level from the ground. This task does not represent realistic scenarios like fast-paced shooting and open-world exploration games that require locomotion with pin-point accuracy and varying elevation levels - which are common in VR. Moreover, their evaluations are based on the movement time (i.e., task completion time) and fail to accommodate crucial performance metrics recommended by Fitts' law, such as error rate and standard deviation of the teleportation selections. 
\update{Furthermore, researchers revealed that standing is preferred for many VR tasks \cite{sarupuri2020testbed, clifton2020effects, coomer2018virtual} due to higher scores and flexibility while sitting offers more safety and precision \cite{zielasko2021sit, zielasko2020sitting}.}
\remove{Furthermore} \update{However}, prior work did not explore the effect of posture on teleportation performance as they \cite{zielasko2020can, zielasko2020either, zielasko2021sit, zielasko2020sitting} were primarily qualitative (using online surveys) and did not investigate user performance of teleportation tasks.
\update{As postures in VR significantly influence overall user performance, it warrants further investigation of its effects within the VR teleportation context.}
Although Fitts' law provides a guideline for target properties, such as target width and distance, no prior work previously explored Fitts' law model for teleportation task in VR. 
In addition, existing research on bi-manual techniques for VR teleportation has \update{often} overlooked the intricacies of tasks like teleporting to targets at varied elevations or to targets with specific width and distance combinations, which are recommended for evaluating human motor performance and mirroring real-world scenarios.

To address the issues of previous research, in this paper, we performed a comprehensive exploration of uni-manual and bi-manual teleportation techniques with Fitts' law analysis for two user postures, sitting and standing, in VR. We first design a set of uni-manual and bi-manual techniques for teleportation tasks for sitting and standing conditions. Additionally, we introduced a modified Fitts' law model specifically adapted for VR teleportation, incorporating factors like the distance from the user and the target's elevation. Subsequently, we conducted a user study involving 20 participants to evaluate these techniques. The tasks involved navigating a VR environment with varying distances, directions, and heights.
Our findings indicate that users are faster and more accurate with bi-manual techniques than uni-manual ones.
We did not find any significant effect of user posture on movement time, error rate, or subjective feedback ratings. 
Furthermore,  results showed that our proposed model consistently outperforms the three other models based on \remove{Adj
$R^2$, AIC, and BIC} \update{the Adjusted $R^2$ (Adj $R^2$), Akaike Information Criterion (AIC), and Bayesian Information Criterion (BIC)} scores.   

This research presents the following contributions:
\begin{itemize}
\item An in-depth exploration of uni-manual and bi-manual teleportation techniques in VR. 
\item An evaluation of users' posture, i.e., sitting vs. standing, for teleportation considering different target widths, distances, and elevations.  
\item A new Fitts' law variation that outperforms existing Fitts' law models for a realistic VR teleportation.
\end{itemize}

%% file: 2_background.tex
\section{Background and Related Work}



Our study extends previous research on teleportation, hands-only locomotion, and bi-manual techniques. With the rise in popularity of head-mounted displays \remove{\cite{bezmalinovic2022cambria, vive2022}} featuring hands-only controls, it is crucial to evaluate this input method for hands-only teleportation. We particularly focus on the parabolic pointer teleportation method, widely used in many commercial controller-based applications, \update{default pointing method in many VR development kits} and investigated by researchers \cite{boletsis2019vr, cherni2020literature, folmer2021teleportation}. In this section, we provide an overview of the existing literature in these intertwined areas.

\subsection{Teleportation}
Within VR locomotion, past studies have primarily researched two main categories: continuous and instant locomotion. Continuous methods simulate real-world actions like walking or flying, while instant methods, like teleportation, allow users to jump directly to their destination. \update{Bowman et al. \cite{bowman1997travel, bowman20043d} categorized travel techniques based on direction/target selection, velocity/acceleration selection, and input conditions and presented a framework for evaluation.}
\update{Walking-in-place (WIP) requires users to provide step-like motions while remaining stationary \cite{usoh1999walking}. 
 Another group of techniques relies on general body motions, such as leaning \cite{marchal2011joyman, tregillus2017handsfree, wang2012comparing}, to navigate in the direction of the body’s center of gravity. 
Griffin et al. \cite{griffin2018evaluation} compared handsfree (walking-in-place and tilt) to handsbusy locomotion techniques (full locomotion and teleportation) and found that handsfree techniques have higher cognitive load and physical demands while offering similar performance.
}


\remove{Teleportation in VR was introduced with} \update{Bolte et al. introduced} "The Jumper Metaphor" \cite{bolte2011jumper}, automatically teleporting users based on their gaze after a timeout. The subsequent "Point and Teleport" \cite{bozgeyikli2016b} technique enabled users to select destinations by aiming at points using a linear pointer.
When compared with prevalent VR locomotion techniques like \remove{"Walking in Place" \cite{usoh1999walking}} \update{WIP} and joystick controls, teleportation was found to be superior in reducing VR sickness, enhancing user perception, and minimizing fatigue \remove{\cite{bozgeyikli2016a, bozgeyikli2016b}} \update{\cite{bozgeyikli2016a, bozgeyikli2016b, weissker2018spatial, zielasko2023stay}}. Its widespread adoption is due to its enjoyable experience, ease of use, and minimal VR sickness \cite{frommel2017effects, matviienko2022skyport} while preserving spatial awareness. The evolution from the straight-line pointer of "Point and Teleport" to the curved parabolic pointer has set a new standard in VR applications \remove{\cite{funk2019assessing, matviienko2022skyport}} \update{\cite{funk2019assessing, chowdhury2022wriarm, chowdhury2023paws}}.
\update{The curvature of the parabolic pointer allows it to land on top of elevated destinations (e.g., staircase) without visual occlusion and aids users in intuitively understanding the VR landscape \cite{coomer2018evaluating}. Although straight-line pointers have been shown to have advantages for mid-air targets \cite{matviienko2022skyport}, we argue that linear pointing might be ineffective in virtual reality scenarios that include elevation.} 
\remove {Willich et al. \cite{von2020podoportation} investigated foot-based controls for the teleportation pointer, highlighting the importance of controller-free teleportation for natural VR interactions. 
The parabolic pointer aids users in intuitively understanding the VR landscape \cite{coomer2018evaluating} and is now common in many VR development kits \cite{mrtk22022, oculus2022, valve2022b}.}

\subsection{Controller-Free Locomotion}
\update{Several studies have explored alternatives to controller-based teleportation, such as jumping forward \cite{bolte2011jumper}, pointing with hand gestures \cite{bozgeyikli2016a, schafer2021controlling}, torso-directed travel \cite{zielasko2020take}, using eye blink or wink \cite{rebsdorf2023blink}. Previous work suggested that these alternatives perform reasonably well compared to the controller-based interfaces \cite{franzluebbers2023versatile, zielasko2016evaluation, prithul2022evaluation} and offer higher spatial awareness and lower simulator sickness \cite{lai2021cognitive}.}

Recent research has shifted towards controller-free teleportation, leveraging hand-tracking technology. Pioneering work by Mine et al. \cite{mine1995virtual} introduced hand-based techniques to manipulate movement direction, emphasizing the effectiveness of using hand movements for natural VR navigation. 
Chastine et al. \cite{chastine2013study} investigated locomotion methods based on hand distances from a leap motion device and found that users easily adapted to controller-free methods, even if hardware techniques were more precise. 
Ferracani et al. \cite{ferracani2016locomotion} examined four hands-based techniques for teleportation and found that the "Tap" technique was preferred based on its precision, comfort, and immersive quality.

\remove{While most instant locomotion research has centered on controller-based applications \cite{cherni2020literature, folmer2021teleportation}, hands-based methods remain relatively unexplored.} 
Bozgeyikli et al. \cite{bozgeyikli2016b} in their research "Point and Teleport" introduced a technique where users direct their index finger toward their desired destination. They showed that "Point and Teleport" was more intuitive than traditional controller-based methods. \remove{Schafer} \update{Sch{\"a}fer} et al. \cite{schafer2021controlling} evaluated uni-manual, bi-manual, and dwell-based teleportation. Their findings demonstrated that hand-based teleportation is highly usable and user-friendly. Chowdhury et al. \cite{chowdhury2022wriarm} designed a technique named "WriArm"  that combined wrist and arm movements for hands-based teleportation, resulting in better performance than arm-based methods. Building on their established merits, we focus on hands-based teleportation in our work.

\subsection{Sitting vs Standing}
The drive to provide immersive VR experiences has led to the exploration of different user postures, particularly the choice between sitting and standing. Researchers have tried to optimize user experiences and address the challenges associated with each posture. The choice of user posture in VR has been found to have a significant impact on the overall user experience, including comfort, immersion, and locomotion.  

Several researchers have analyzed the influence of posture, specifically sitting and standing, on 3D interaction in VR environments in terms of cyber-sickness, comfort, precision, safety, vertigo, engagement, complexity, flexibility, and accessibility \cite{zielasko2020sitting}. Zielasko et al. \cite{zielasko2021sit} have found that sitting posture provides more safety, comfort, precision, and accessibility and reduces cybersickness, whereas Standing enhances engagement levels. However, the performance and user preferences vary with the specific tasks performed in VR environments. Studies suggest that both postures are almost equally preferred, with participants reporting more comfort when seated but more natural, intuitive and immersive interaction when standing in VR space exploration context \cite{coomer2018virtual} while standing is more preferred due to improved task collection scores, flexibility and immersion but leads to more cybersickness in VR locomotion context \cite{sarupuri2020testbed, clifton2020effects}. According to Leyrer et al., \cite{leyrer2011influence, leyrer2015eye}, standing for locomotion is more natural as eye height affects the distance estimation in VR. Prior work has suggested that there is no clear winner between sitting and standing in VR environments, but it depends on the mismatch between the real-world physical posture and virtual posture in the VR task \cite{zielasko2020can, zielasko2020either}. Meanwhile, users have been reported to have a tendency to physically sit while exploring the VR space \cite{digitaltrends_vr}.  

We are motivated by these works to study the impact of posture on hands-only VR teleportation.

\subsection{Bi-Manual Techniques}

Hands-based VR locomotion techniques have evolved significantly, with recent studies focusing on the advantages of bi-manual techniques for navigation and interaction in VR environments. These studies have laid the groundwork for understanding the benefits of using both hands, not only for locomotion but also for creating more immersive and intuitive user experiences in VR. 
\update{Nancel et al. \cite{nancel2011mid} compared between bi-manual and uni-manual mid-air target pointing and concluded that bi-manual techniques consistently outperform uni-manual techniques.}
Previous work has explored the potential of bi-manual input modalities for 3D interaction in VR environments \cite{caputo2017single, lubos20144}. They have found bi-manual interactions to be advantageous over uni-manual interactions in a confined interaction space, making interaction more efficient in certain scenarios. 
Recent studies have evaluated the efficiency, usability, perceived workload, and user preference of bi-manual Techniques for Locomotion in VR \cite{schafer2022controlling, zhang2017double}. Zhang et al. \cite{zhang2017double} have compared a bi-manual technique, “DHGI (Double Hand Gesture Interaction)”, to joystick and portal methods \cite{matviienko2022skyport} and found it to be more intuitive and immersive and to reduce motion-sickness. They found DHGI to be better than uni-manual gestures as users had limited interaction potential while using only one hand.   
\remove{Schafer} \update{Sch{\"a}fer} et al. \cite{schafer2021controlling} did a comparative evaluation of bi-manual and uni-manual Techniques in the context of VR Teleportation. Specifically, their research explored whether two hands provide better control or if a single hand is sufficient. They concluded that there is no clear winner between uni-manual and bi-manual techniques in terms of both task performance and user preference. Some users preferred bi-manual techniques due to its higher control than a uni-manual technique, while some users reported the uni-manual technique to be more comfortable and even perceived as faster.

\subsection{Summary}
Teleportation is a widely used VR locomotion technique due to its \remove{reduced} \update{lower} VR sickness \update{impact}, improved user satisfaction, and lower fatigue \cite{chowdhury2022wriarm, bozgeyikli2016a, bozgeyikli2016b, frommel2017effects, matviienko2022skyport}. 
Although researchers \cite{zielasko2020can, zielasko2020either, zielasko2021sit, zielasko2020sitting} have extensively explored the effect of posture on teleportation performance based on qualitative surveys, they did not investigate actual user performance in teleportation tasks.
Furthermore, we found only one previous research \cite{schafer2021controlling} that compared bi-manual and uni-manual teleportation techniques and found no significant performance difference between them. However, they only considered simple tasks with identical large targets placed uniformly on the same elevation level at equal distances from each other, which doesn't represent realistic scenarios in VR applications. Furthermore, no previous research on teleportation used a standardized procedure or model such as the Fitts' law to evaluate user performance.

%% file: 3_design-space.tex
\section{Teleportation Techniques}
\update{We focus on evaluating bi-manual techniques for hand-based teleportation while using well-explored uni-manual techniques as the baseline, following prior work \cite{chowdhury2022wriarm, chowdhury2023paws, schafer2021controlling}.} 
Based on prior work, we observed that  the teleportation tasks are divided into two subtasks: \textbf{Pointer Control} and \textbf{Selection}.\\
    \textbf{Pointer Control}: In this sub-task, users are required to control the teleportation parabola \remove{(also known as the ray)} with their hand to point to a destination. We used a similar approach as in \cite{schafer2021controlling, chowdhury2022wriarm}, where \update{the teleportation parabola originated from the users' hand and the casting direction was based primarily on the orientation of the arm and wrist. } Users' arm movements were utilized for coarse movements and wrist movements such as flexion/extension and radial/ulnar deviation were used to control the \remove{fine movement} \update{launch angle and fine movements, respectively.
    The distance traveled by the parabola was manipulated by retracting or extending the arm. We use Kalman filters to increase the pointer stability and add a last-second spike compensation to avoid unintended \remove{trigger movements}\update{confirmation}, similar to previous work \cite{chowdhury2023paws, wolf2020understanding}.}\\\textbf{Selection}: In this subtask, users are required to perform a hand gesture to \remove{trigger}\update{confirm} a selection and teleport to a destination where the end of the teleportation parabola is pointing. Based on \cite{chowdhury2022wriarm, schafer2021controlling}, we used 
    `closing the index finger' gesture to confirm a selection. \update{Our source code is available at this \href{https://github.com/SiddhanthRaja/Exploring-Bi-Manual-Teleportation-in-Virtual-Reality}{\textit{GitHub link}}}.

A user can control the \remove{ray} \update{teleportation parabola} with any of their hand and \remove{trigger}\update{confirm} a selection with the same hand or the other hand. Based on this, we identified four combinations. In addition, we introduced a Dwell-based selection method as this technique was used in previous work \cite{schafer2021controlling, matviienko2022skyport, bozgeyikli2016b}.

    \begin{figure}[h!]
	\centering
	\includegraphics[width=1.0\columnwidth]{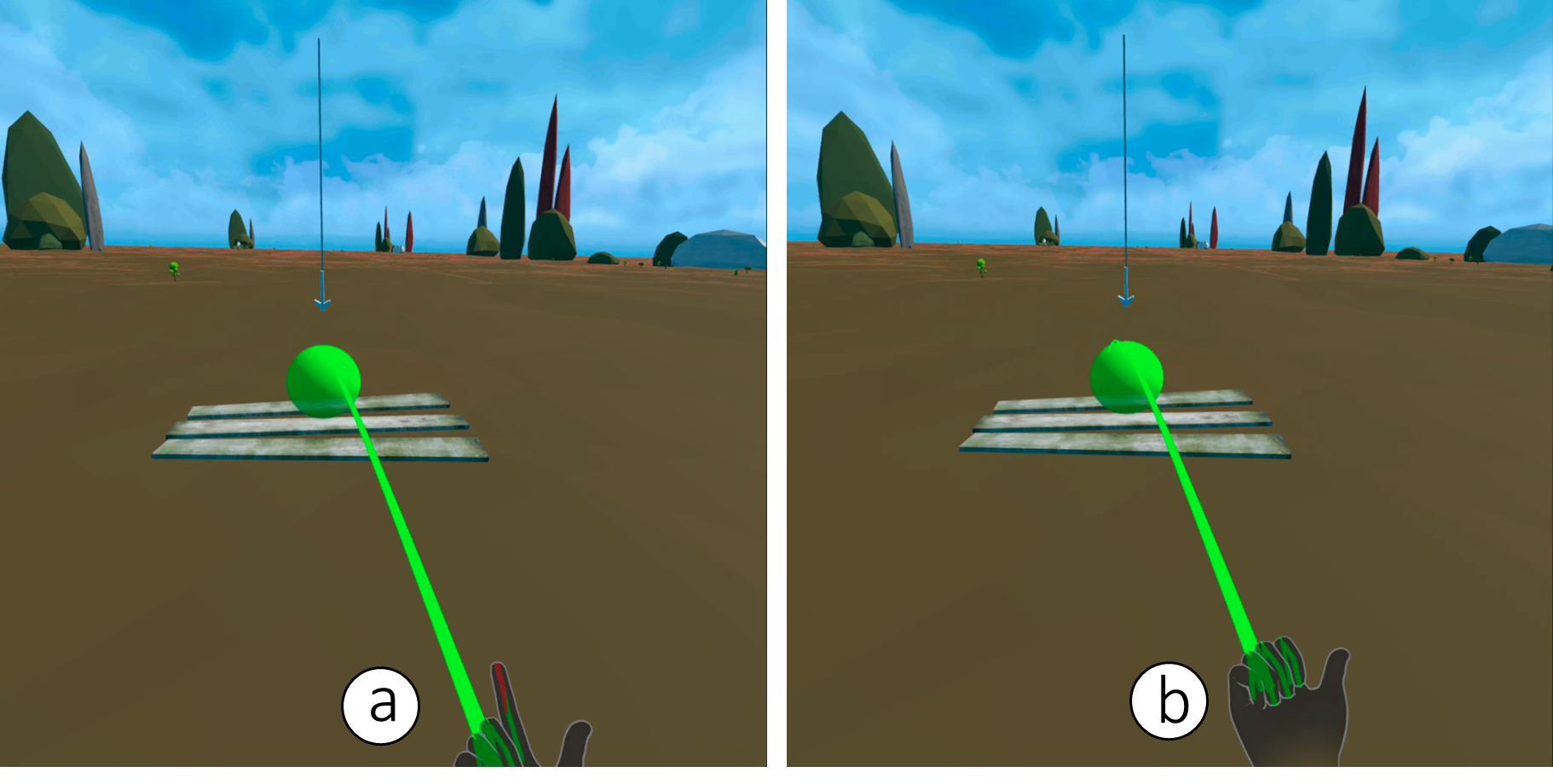}
	\caption{Right Pointer, Right Gesture (RPRG). (a) A user controls the teleportation parabola using the right hand, and (b) selection is done by closing the index finger of the same hand.} 
	\label{fig:rprg} 
\end{figure}

\begin{itemize}
    \item \textbf{Right Pointer, Right Gesture (RPRG)}: In this technique, the \remove{ray} \update{teleportation parabola} is controlled with the right hand, and the same hand is used to \remove{trigger}\update{confirm} a selection (see Figure \ref{fig:rprg}). One of the advantages of this technique is that users can control the pointer and \remove{trigger}\update{confirm} selections using the same hand, thus minimizing any switching costs and reducing the teleportation time. It also reduces users' cognitive load, as they do not need to mentally switch between hands for different actions and focus on one hand for pointer control and selection. However, we anticipate this technique will lead to more errors, as the same hand is used for pointer control and selection. More specifically, users might accidentally \remove{trigger}\update{confirm} selections while intending to point, leading to unintended errors. \update{This is known as the Heisenberg effect of spatial interaction \cite{wolf2020understanding}}

               \begin{figure}[h!]
    \centering
    \includegraphics[width=1.0\columnwidth]{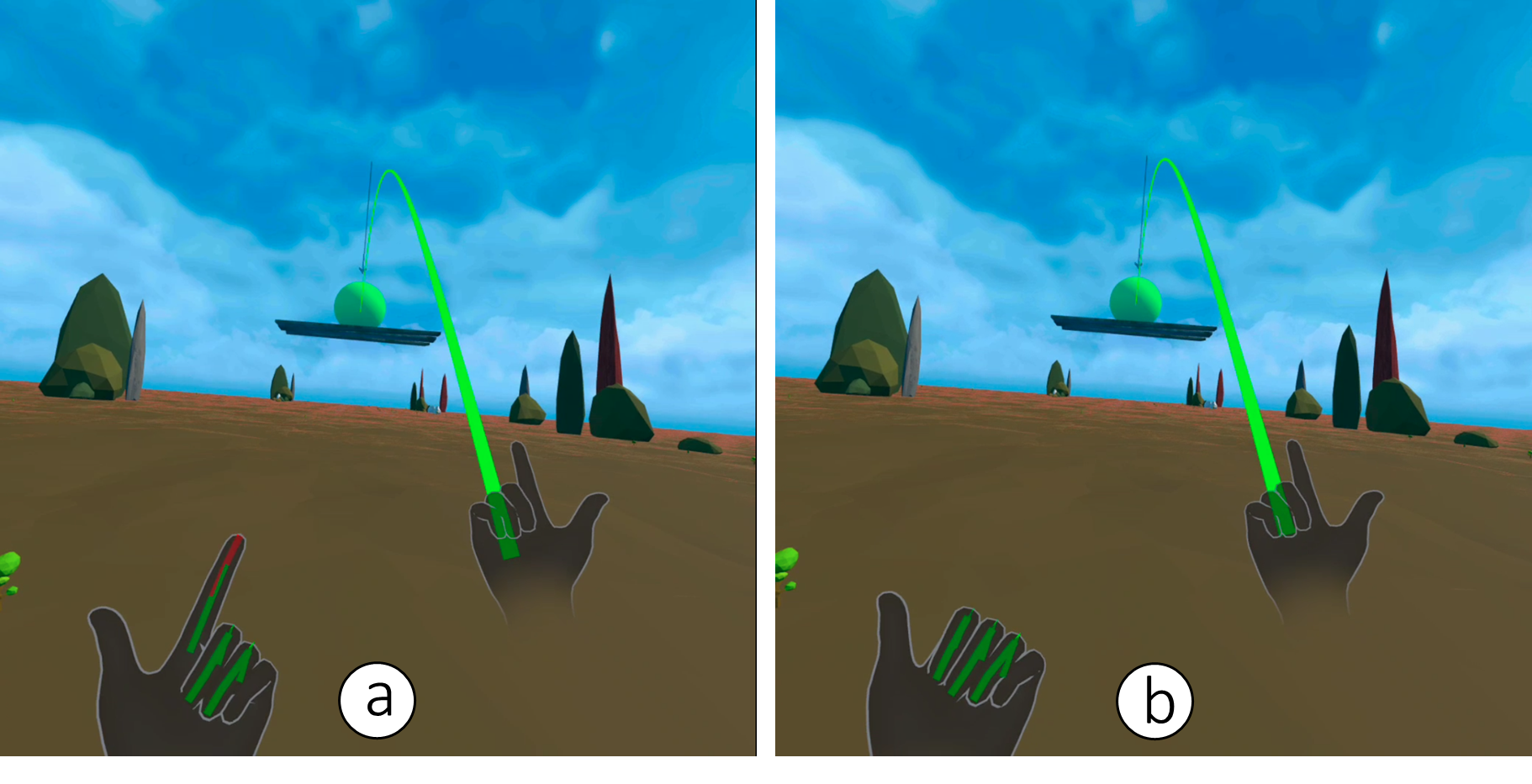}
	\caption{Right Pointer, Left Gesture (RPLG). (a) A user controls the teleportation parabola using the right hand, and (b) selection is done by closing the index finger of the left hand.} 
	\label{fig:rplg} 
\end{figure}

    \item \textbf{Right Pointer, Left Gesture (RPLG)}: 
    \remove{In uni-manual techniques such as RPRG, the same hand is used to control the pointer and confirm selections, leading to sudden hand movements during selection (e.g., closing the index finger) that can result in erroneous selections. Although we've implemented software-based solutions for stabilization, such as handling jitters and smoothing using Kalman filters, sudden hand movements during the confirmation gesture can still impact performance. Therefore, in this technique, users can control the pointer and trigger selections using different hands, offering more control and stability.} 
    With RPLG, the \remove{ray} \update{teleportation parabola} is controlled with the right hand, and the left hand is used to make a gesture to \remove{trigger}\update{confirm} a selection (see Figure \ref{fig:rplg}). By involving two hands for tasks requiring coordination and control, we anticipate RPLG to have a steeper learning curve than other techniques. However, once learned, we expect this technique to outperform uni-manual techniques \update{as users can control the pointer and \remove{trigger}\update{confirm} selections using different hands, offering more control and stability.}

    \begin{figure}[h!]
    \centering
    \includegraphics[width=1.0\columnwidth]{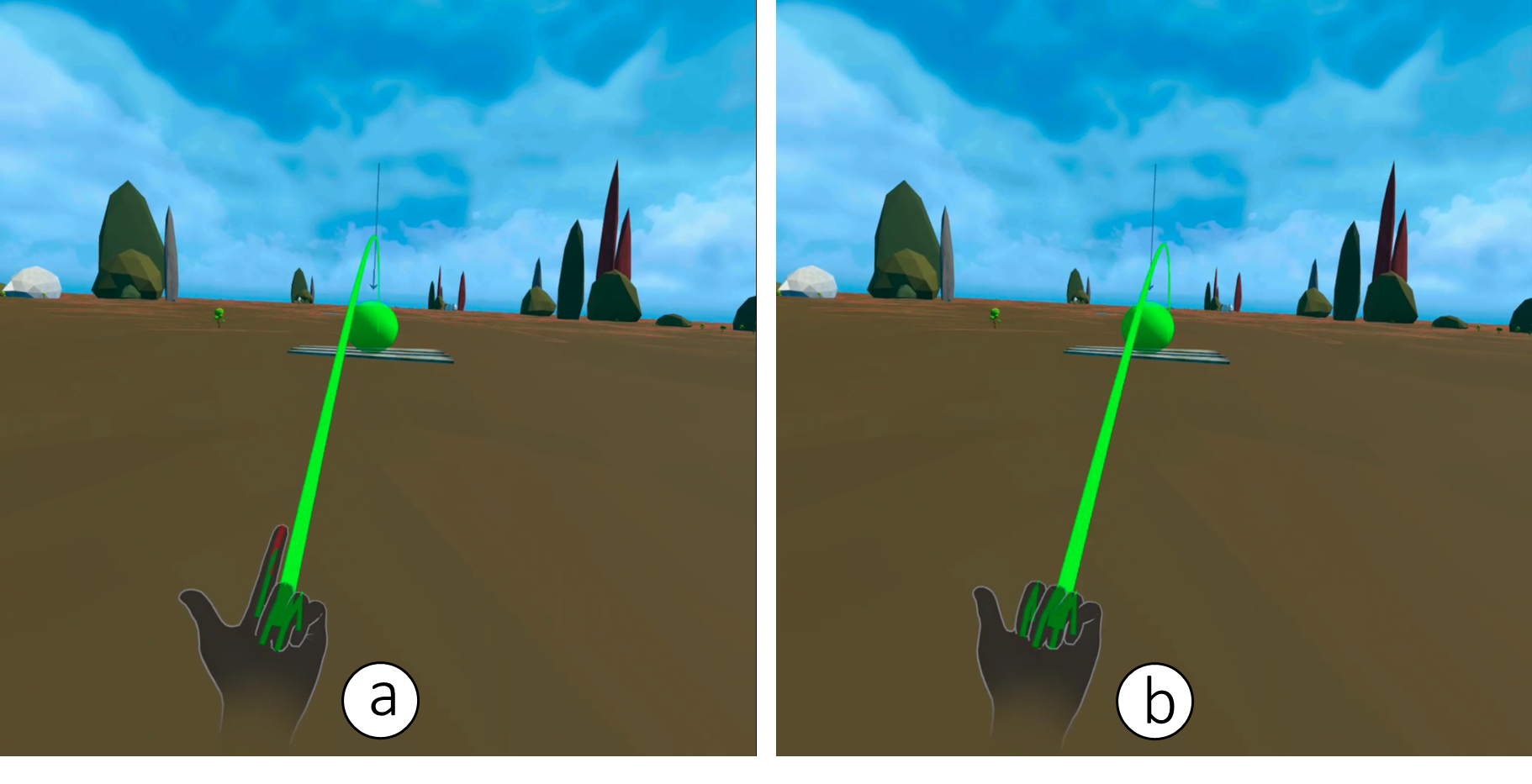}
	\caption{Left Pointer, Left Gesture (LPRG). (a) A user controls the teleportation parabola using the left hand, and (b) selection is done by closing the index finger of the same hand.} 
	\label{fig:lplg} 
\end{figure}

    \item \textbf{Left Pointer, Left Gesture (LPLG)}: With LPLG, the \remove{ray} \update{teleportation parabola} is controlled with the left hand, and the same hand is used to \remove{trigger}\update{confirm} a selection (see Figure \ref{fig:lplg}). This technique shares similar advantages with RPRG \remove{, such as lower switching costs, reduced learning time, and decreased cognitive load for users since the same hand is involved in both pointer control and selection}. However, we anticipate that LPLG will be slower than RPRG because, with this technique, \remove{users} \update{participants (exclusively right-handed in our study)} are using their non-dominant hand to control the \remove{ray} \update{teleportation parabola} and \remove{trigger}\update{confirm} the selection, which could result in longer teleportation times and higher error rates compared to RPRG.
    
       \begin{figure}[h!]
    \centering
    \includegraphics[width=1.0\columnwidth]{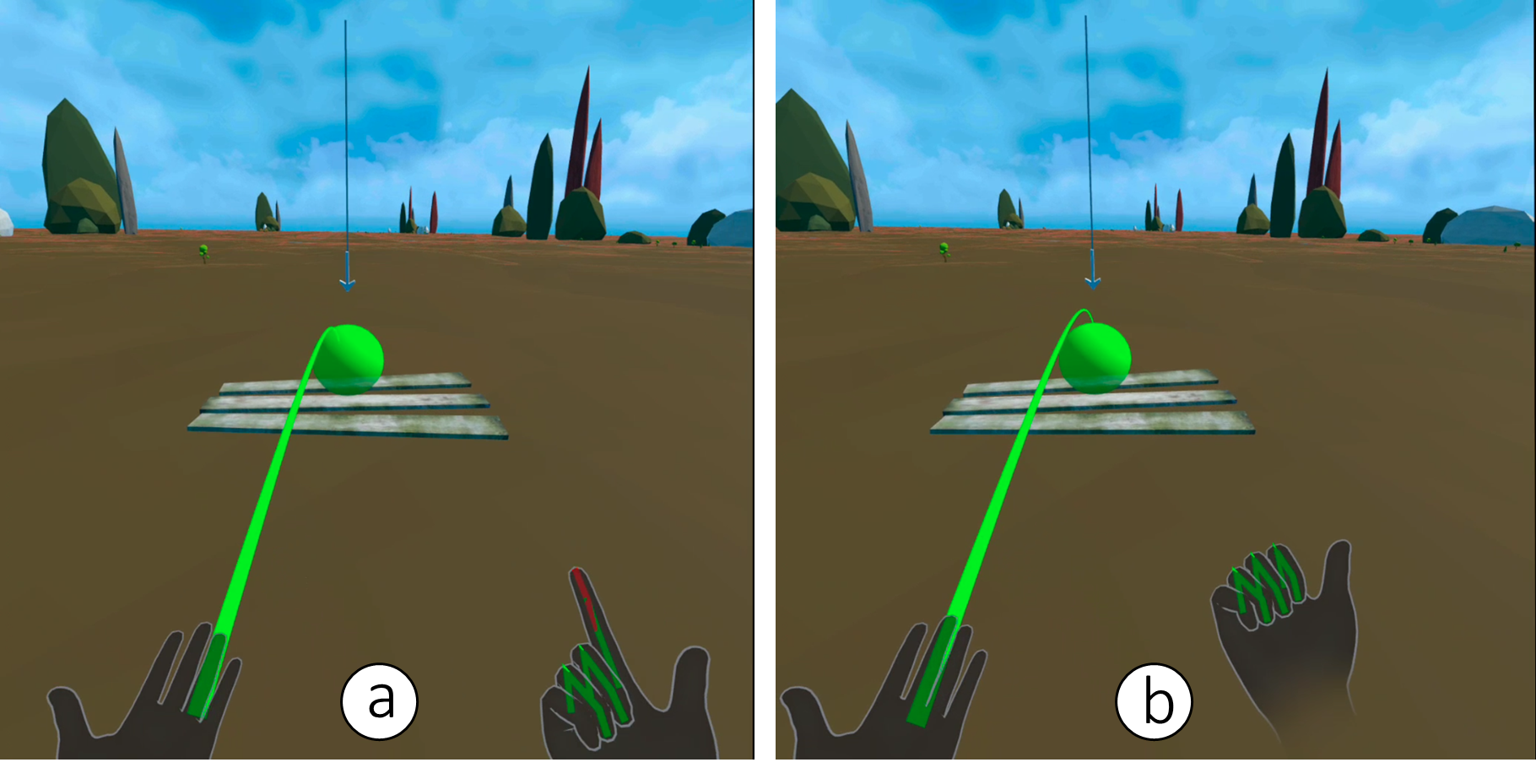}
	\caption{Left Pointer, Right Gesture (LPRG). (a) A user controls the teleportation parabola using the left hand, and (b) selection is done by closing the index finger of the right hand.} 
	\label{fig:lprg} 
\end{figure}

    \item \textbf{Left Pointer, Right Gesture (LPRG)}: In this technique, the \remove{ray} \update{teleportation parabola} is controlled with the left hand, and the right hand is used to make a `closing index finger' gesture to \remove{trigger}\update{confirm} a selection (see Figure \ref{fig:lprg}). This technique shares the same advantages as RPLG since both hands have specific tasks. However, we anticipate its performance to be lower than RPLG because the non-dominant hand is responsible for pointer control, demanding precise movements to target accurately. Nonetheless, it is expected to outperform techniques where the non-dominant hand handles both teleportation and selection.

   \begin{figure}[h]
    \centering
    \includegraphics[width=1.0\columnwidth]{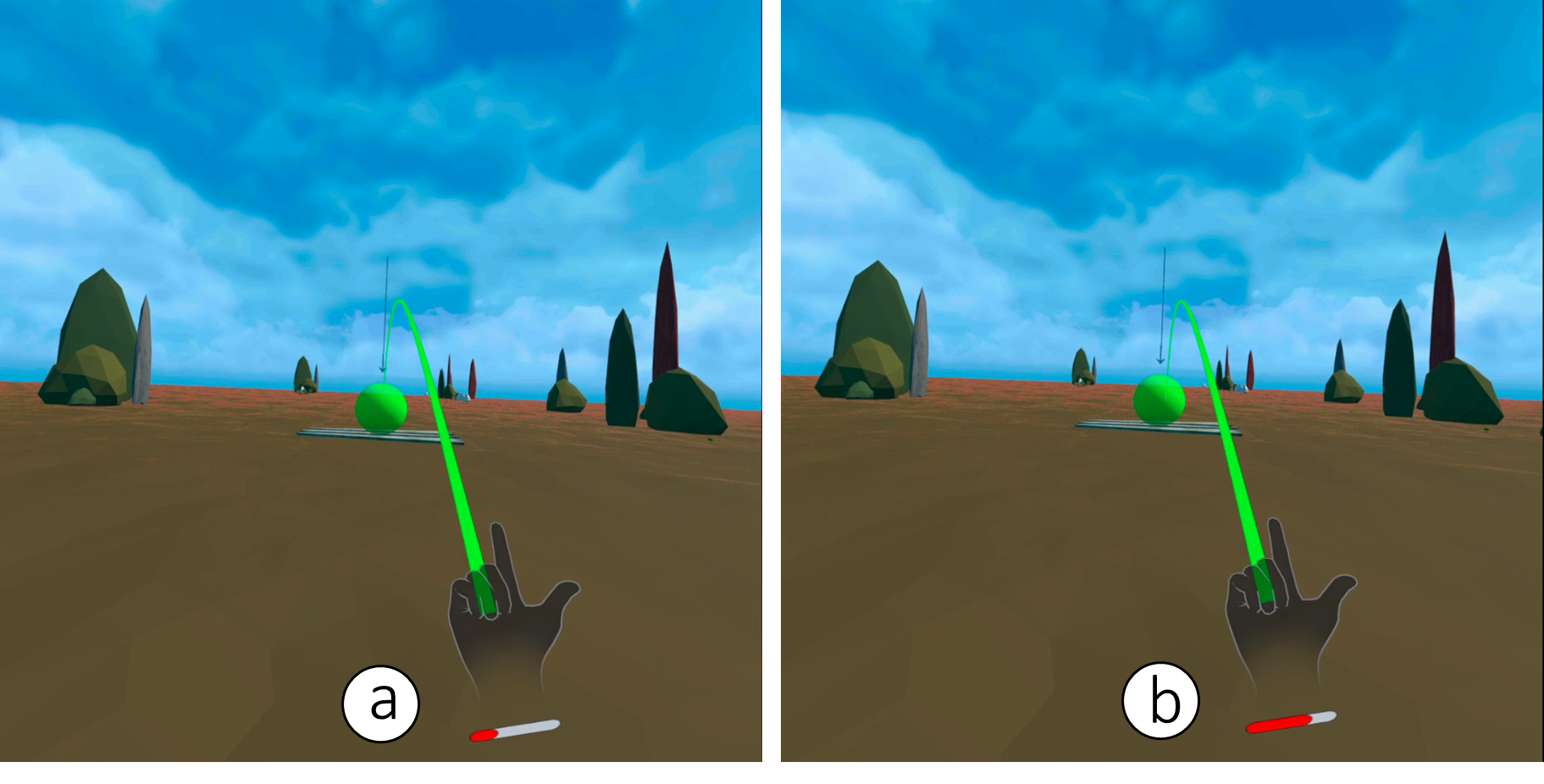}
	\caption{Right Pointer with Dwell (LPRG). (a) A user controls the teleportation parabola using the right hand, and (b) selection is done by holding the arm static until timeout (indicated by the progress bar over the wrist)} 
	\label{fig:rpdw} 
\end{figure}

    \item \textbf{Right Pointer with Dwell (RPDW)}: The \remove{ray} \update{teleportation parabola} is controlled with the right hand in this technique. Unlike other techniques, the `closing index finger' is not used for \remove{trigger}\update{confirm}ing a selection. Instead, users were required to keep their arm static within a 0.3 m radius for a certain threshold (see Figure \ref{fig:rpdw}). For this study, we used an 800ms threshold as a similar dwell time was used in \cite{namnakani2023comparing} and identified to be the most effective dwell time in VR contexts. A selection is \remove{trigger}\update{confirm}ed automatically after reaching the threshold. To enhance user experience and reduce potential frustration, we added a progress bar over the wrist to provide feedback on the dwell time (see Figure \ref{fig:rpdw}). In uni-manual techniques, performing a confirmation gesture (closing the index finger) can lead to sudden hand movements, potentially resulting in erroneous selections. Additionally, holding the arms in the air and repeatedly extending and flexing the fingers may cause fatigue. Therefore, we anticipate that this technique will offer more stability and comfort to the participants. However, the timeout duration might lead to frustration for easily reachable targets due to a long wait time, while also imposing temporal demands.

\end{itemize}

\section{Fitts' Law variations}

Fitts' law, initially developed as a predictive model for human motor behavior in 1D pointing tasks \cite{fitts1954information}, has been subsequently extended to consider 2D \cite{mackenzie1992fitts} and 3D pointing selection tasks \cite{triantafyllidis2021challenges}. 
Since teleportation intrinsically requires pointing at targets at different locations in a virtual environment, utilizing a human motor movement model such as Fitts' law is crucial for its evaluation.
Triantafyllidis and Li \cite{triantafyllidis2021challenges} pointed out the need for a standardized human performance metric in 3D spaces as existing models have primarily focused on a single input or environment scenario and have not been evaluated in real-life pointing scenarios (such as Teleportation tasks). Therefore, developing a model specifically tailored for teleportation in VR is necessary. Before delving into our proposed model, we explore the Fitts' law model and its extensions used in different contexts.

\subsection{One-part Model} 
The widely used form of Fitts' law for pointing tasks, known as the Shannon-Fitts' formula or ISO 9241-9 \cite{iso20009241}, was proposed by MacKenzie \cite{mackenzie1992fitts}. This formula enables the prediction of pointing movement time (MT) using the equation:
\begin{equation} \label{eqn_Fittslaw_popular}
MT = a + b \log_2\left(\frac{A}{W} + 1\right)
\end{equation}
In the equation, A represents the movement amplitude, W represents the target width, and the constants a and b are empirically determined.
The Index of Difficulty (ID) of a target corresponds to the logarithmic term $\log_2\left(\frac{A}{W} + 1\right)$.

Previous research has proposed several extensions to account for different conditions and improve the modeling of Fitts' law. These extensions include considering target height \cite{crossman1956measurement}, incorporating movement directions in degrees \cite{kopper2010human}, or addressing 3D environments \cite{grossman2004pointing}, among others.

In the following sections, we describe two variations of Fitts' law that are conceptually relevant to our use case of virtual pointing with head-mounted displays (HMDs). These variations are the ``Two-part model'' \cite{janzen2016modeling, shoemaker2012two} and the ``Vergence-Accomodation Model'' \cite{barrera2019effect}.

\subsection{Two-part Model}

Kopper et al. \cite{kopper2010human} conducted a study demonstrating that the angular width and angular amplitude of a target significantly influence the difficulty of pointing tasks. However, Shoemaker et al. \cite{shoemaker2012two} and Janzen et al. \cite{janzen2016modeling} proposed a variation of Fitts' law called the ``two-part model'' to consider the effects of target width and target amplitude separately.

They found that the two-part model, also referred to as the Shannon-Welford formula, provided a better fit than using angular measurements alone. The formula for the two-part model, based on the Welford version of Fitts' law, predicts movement time (MT) using the equation:

\begin{equation} \label{eqn_Fittslaw_twopart}
MT = a + b_1 \log_2(A + W) - b_2 \log_2(W) 
\end{equation}

In the equation, A represents the target amplitude, W represents the target width, and k is a constant that is empirically determined based on factors such as cursor gain and user distance from the display. The constants a and b are derived through regression analysis.

Since the two-part model demonstrates a better fit than using angular measurements, we consider this model in our analysis.

\subsection{Vergence-Accomodation Model}

In VR displays, perceptual conflicts such as the vergence-accommodation conflict \cite{hoffman2008vergence} are known to negatively impact users' ability to localize targets in 3D space.
Recent work by Machuca and Stuerzlinger \cite{barrera2019effect} found that when performing reciprocal selection tasks for 3D targets, participants were significantly slower to select virtual targets presented via a screen-based stereoscopic display than physical targets. They proposed a Fitts' law variation for stereoscopic displays: 
 \begin{equation} \label{eq:FittsID_Stereo}
M T=a+b\log _{2}\left(\frac{A}{W}+1\right)+c(CTD)  
\end{equation}
 where CTD is the change in target depth between the current target and the previous target. A represents the movement amplitude, W is the target width, and a, b, and c are empirically evaluated constants.

\subsection{Proposed Model}

Although the Vergence model addresses a key drawback of prior research by incorporating the change in depth between two targets, the model has some limitations. The first is that the index of difficulty of a pointing task should not have any unit. Machuca and Stuerzlinger \cite{barrera2019effect} recognized that incorporating CTD (in centimeters) introduces a unit term, which is not ideal as the original Fitts' law \cite{fitts1954information} and its variations define index of difficulty of a task to be unitless. Moreover, previous research has found that target depth \cite{kopper2010human} significantly affects pointing performance due to vision parallax \cite{wagner2023fitts}. Additionally, if a target is positioned at a higher altitude but at a lower depth, the disparity in height will have a more significant impact on movement time compared to the target's lower depth. Based on these, we propose a new model to solve the Vergence-accommodation conflict: 

\begin{equation} \label{eqn_Fittslaw_proposed}
MT = a + b_1\log _{2}\left(\frac{A}{W}+1\right) - b_2\log _{2}\left(\frac{W}{MAX(D,H)}+1\right)
\end{equation}
 where D is the distance in the depth direction from the user to the target, and H is the difference in altitude from the user to the target. A represents the movement amplitude, W is the target width, and a, b1, and b2 are empirically evaluated constants. As we hypothesize high values of either the depth-distance or the altitude will affect the movement time, we consider the highest values between these two factors for our formula.

%% file: 5_exp-protocol.tex
\section{User Study}


We conducted a user study to explore the following research questions: 

\begin{itemize} 
\itemsep0.5em 
\item RQ1: Which hand-only teleportation technique(s) are the most effective for VR teleportation? 

\item RQ2: How does user posture impact performance and preferences using these teleportation techniques?

\item RQ3: Can the proposed variations of Fitts' law model accurately predict users' teleportation performance in VR?
\end{itemize}

\subsection{Participants}
We recruited 20 right-handed participants as we did not want hand dominance to become a confounding factor. Our participants included 13 males and 7 females aged between 20 and 33 (mean age = 23.8, SD = 4.13). Recruitment was facilitated through on-campus advertisements and word of mouth. Participants were compensated with \$15 for their involvement. Of these, 12 had prior VR experience (i.e., who used VR devices at least once in the past 12 months), and 4 were familiar with controller-based teleportation.

\subsection{Apparatus}

The \remove{Oculus} \update{Meta} Quest Pro VR headset was used for its integrated hand-tracking capabilities, eliminating the need for external tracking solutions. We developed the virtual environment and techniques using Unity 3D (2020.3.33f1) and C\# scripting. The Oculus Integration SDK provided real-time hand-tracking data. The techniques were developed on a computer equipped with an RTX 3070 GPU. Data from the study sessions was logged to a Cloud-Firestore database.
\update{An office swivel chair was used to allow unrestricted turns/movement of participants in seated posture.}

\subsection{Task and Procedure}

\begin{figure}[h!]
    \centering
    \includegraphics[width=1.0\columnwidth]{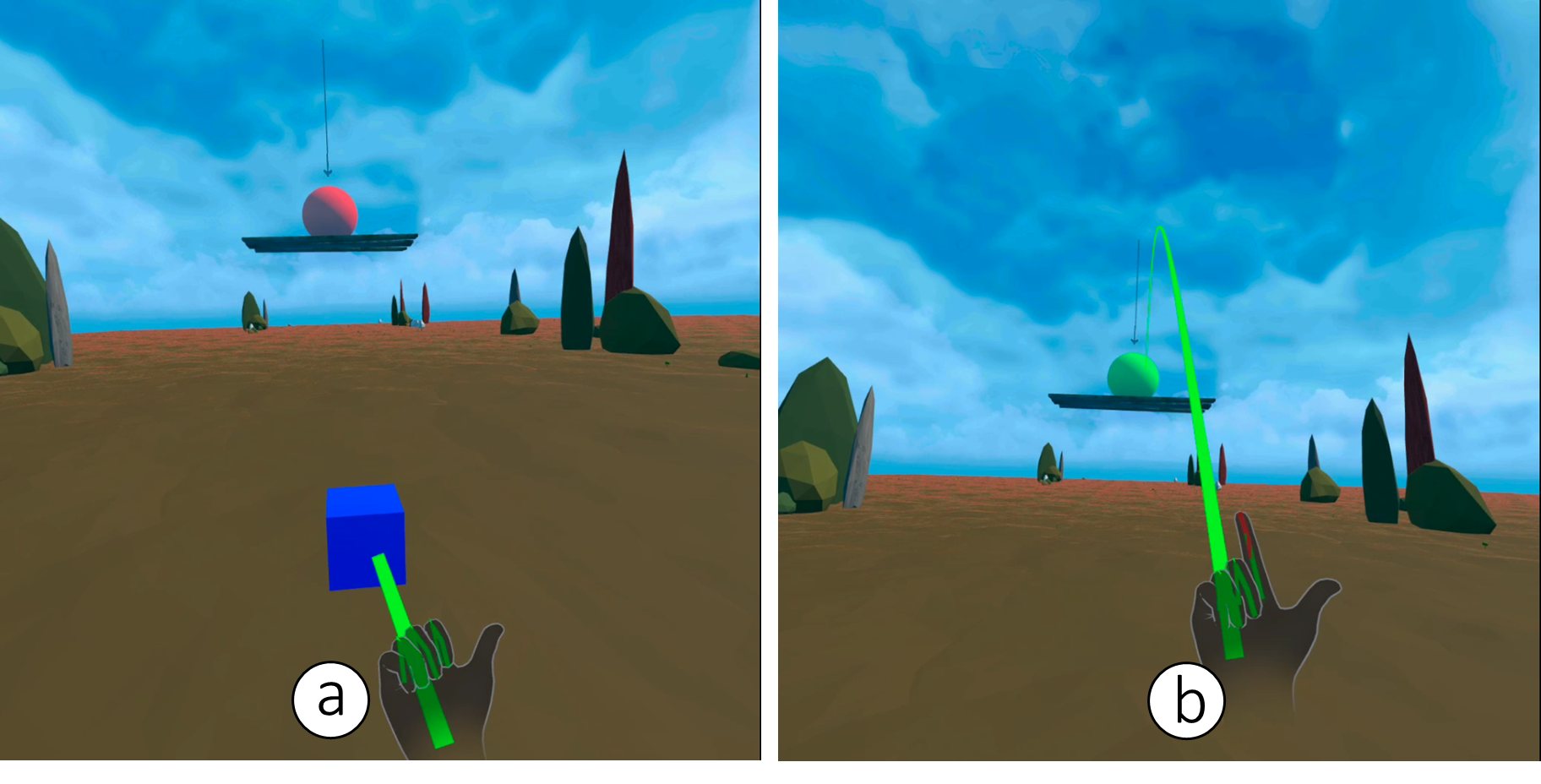}
	\caption{(a) At the beginning of the task, participants were instructed to point to a blue cube and \remove{trigger}\update{confirm} a selection, ensuring that each trial began from a fixed position. (b) Once the blue cube was selected, participants could move the teleportation pointer to a destination using arm and wrist movements.} 
	\label{fig:tasks} 
\end{figure}

Participants were asked to teleport from one location to another designated destination using each technique, as illustrated in Figure \ref{fig:tasks}. 
\update{Based on previous research on 3D pointing \cite{teather2014visual, wagner2023fitts, triantafyllidis2021challenges}, we highlighted the targets when the teleportation parabola intersected with the target to provide visual feedback.}
Teleportation was confined to platforms through a singular move for our controlled evaluation. We used it as similar designs were used in previous work for evaluating teleportation techniques \cite{chowdhury2023paws, chowdhury2022wriarm, funk2019assessing, matviienko2022skyport}. More specifically, we used platforms and single move.
    \textbf{Platforms} provided controlled teleportation distances and elevations. They also offered participants a tangible destination rather than leaving them suspended in mid-air. Platforms emulate terrain-based elevation changes while maintaining the confines of our study conditions.
    \textbf{Single move} constraint ensured the teleportation was consistent, eliminating variables like multiple short-distance moves. This also helps us to allow for broader generalizations, such as multiple teleportation. 

The trial commenced by displaying a blue cube (referred to as the `Start button') and a destination for teleportation. The start button was positioned at a fixed distance of 0.59 m in front of the user's view. Participants were required to select it using the 'closing index finger' gesture to begin a trial. This setup ensured users started each trial from a fixed position (see Figure \ref{fig:tasks}a). Once participants selected the start button, it disappeared, and the destination remained visible. Destinations were indicated by red translucent spheres of varying widths (diameter) placed on 1-meter square platforms. The width of these spheres (referred to as the `target') and the distance from the users varied based on the study condition. An overhead arrow guided participants to the destination platform within the VR environment. The color of the sphere and parabolic pointer changed from red to green when users accurately aimed at the sphere, providing visual feedback (see Figure \ref{fig:tasks}b). Participants confirmed their teleportation using the `closing index finger' gesture. Successful teleportations were followed by an audio cue, and the start and next destinations appeared. Failed attempts were included in the trial sequence, ensuring participants successfully teleported to all designated destinations.

In studies involving Fitts' law, it is customary to vary target width and distance to encompass a wide range of Index of Difficulty (ID) values, typically between 2 bits and 8 bits according to the ISO 9241-9 standard \cite{soukoreff2004towards,iso20009241}. Following this principle, we positioned the targets in different locations, varying the following factors:

    \textbf{Distance:} Building on previous research \cite{chowdhury2022wriarm, chowdhury2023paws, matviienko2022skyport}, we manipulated the distance between the start cube and the target position. Specifically, we examined two teleportation distances: 3m and 9m.
    
    \textbf{Height:}
    Based on prior research \cite{chowdhury2022wriarm, chowdhury2023paws}, we varied the height of the target, defined by the elevation difference between the start and the target, between ground level (0 m) and upper level (3 m) leading to planar (from 0 m to 0 m), upward (from 0 m to 3m), and downward trajectories (from 3 m to 0 m).

    \textbf{Target Width:} We used two target widths based on a pilot study. We defined the target width by the diameters of the target spheres, which were 0.2 m and 1.35 m.
    
    \textbf{Angle:} The target appeared at a viewing angle of -10$\degree$, 0$\degree$ or +10$\degree$ picked randomly to minimize directional bias.
    This angle was chosen to ensure that targets were consistently visible on the screen. This approach helped participants focus on teleportation without the need to visually search for targets located off-screen.

During the study, participants wore the Meta Quest Pro headset. An experimenter (one of the authors) conducted the study, providing participants with an introduction to VR, explaining locomotion and teleportation, and outlining the study's objectives. Participants were then given a set of practice trials (8 in total), which were applied to each technique before the main trials began. The main phase consisted of 10 blocks of \remove{60} \update{40} trials each, with breaks in between to prevent fatigue. During these breaks, participants completed questionnaires to provide NASA TLX and SUS scores. Once participants completed teleportation tasks with all the techniques, they reviewed their responses and participated in a brief interview. Each session lasted approximately 90 minutes.

\subsection{Design}

We used a within-subjects design, focusing on two primary independent variables: \textit{Technique} (RPRG, LPLG, RPLG, LPRG, and RPDW) and body \textit{Posture} (Sitting and Standing).
We used a Latin square design to counterbalance the \textit{Technique-Posture} combinations. The combinations of Target Size, Distance, and Height variations were presented randomly for each \textit{Technique-Posture} condition.
In total each participant performed 5 \textit{Technique} $\times$ 2 \textit{Posture} $\times$ 2 \textit{Size} $\times$ 2 \textit{Distance} $\times$ 2 \textit{Height} variation $\times$ 5 repetitions, totaling 400 trials.
We recorded movement time, error rate, and endpoint deviation from the target center for each trial. 
Movement time was measured from the start of each trial when participants selected the `start' button to their successful selection of the target sphere. 
The error rate was calculated based on the percentage of trials with incorrect selection attempts, where the selection was made outside the target boundary.
Endpoint deviation was determined by measuring the shortest distance between the target sphere's center and the location of the parabolic pointer during a successful selection. 
Additionally, we collected subjective data through the NASA TLX questionnaire.

%% file: 6_exp-results.tex
\subsection{Results}
We analyze movement time and error data with repeated measures ANOVA on log-transformed data and post-hoc pairwise comparisons with Bonferroni corrections.
In sphericity violation, we report Greenhouse-Geisser corrected p-values and degrees of freedom.

\subsubsection{Movement Time}
We found significant main effects for the independent variables \textit{Technique} ($F_{4, 72} = 18.31$, $p < 0.001$, $\eta^2$ = 0.50), \textit{Distance} ($F_{1, 18} = 551.17$, $p < 0.001$, $\eta^2$ = 0.97), \textit{Height} ($F_{1, 18} = 459.43$, $p < 0.001$, $\eta^2$ = 0.96), and \textit{Width} ($F_{1, 18}  = 1267.57$, $p < 0.001$, $\eta^2$ = 0.99). 
However, \textit{Posture} ($F_{1, 18} = 0.53$, $p = 0.47$) did not have a significant effect on movement time. 
The mean movement time is 2.58s (95\% confidence interval, CI: [2.42, 2.74]) for RPRG, 2.41s (CI: [2.25, 2.56]) for RPLG, 2.71s (CI: [2.54, 2.88]) for LPLG, 2.61s (CI: [2.44, 2.77]) for LPRG, and 2.88s (CI: [2.73, 3.02]) for RPDW. 
Post-hoc pairwise comparisons reveal that pointing time with RPLG is significantly faster than the four other \textit{Techniques} (all $p<0.001$).
Conversely, RPDW is significantly slower than the four other \textit{Techniques} (all $p < 0.001$).
Additionally, RPRG is significantly faster ($p<0.05$) than LPLG.
As expected, the higher target \textit{Width} (1.35m (M = 1.67s, [1.63, 1.71]) leads to significantly faster selection ($p<0.001$) than lower target widths (0.2m: M = 3.6s, CI: [3.51, 3.7]).  
The lower target \textit{Distance} (3m, M = 2.03s, CI:[1.96, 2.1]) leads to significantly faster selection ($p<0.001$) than higher target \textit{Distance} (9m, M = 3.24s, CI:[3.13, 3.35]).
Similarly, lower target \textit{Height} (0m, M = 2.27s, CI:[2.18, 2.36]) leads to significantly faster selection ($p<0.001$) than higher target \textit{Height} (3m, M = 3.00s, CI:[2.9, 3.11]).

  \begin{figure}[h!]
    \centering
    \includegraphics[width=1.0\columnwidth]{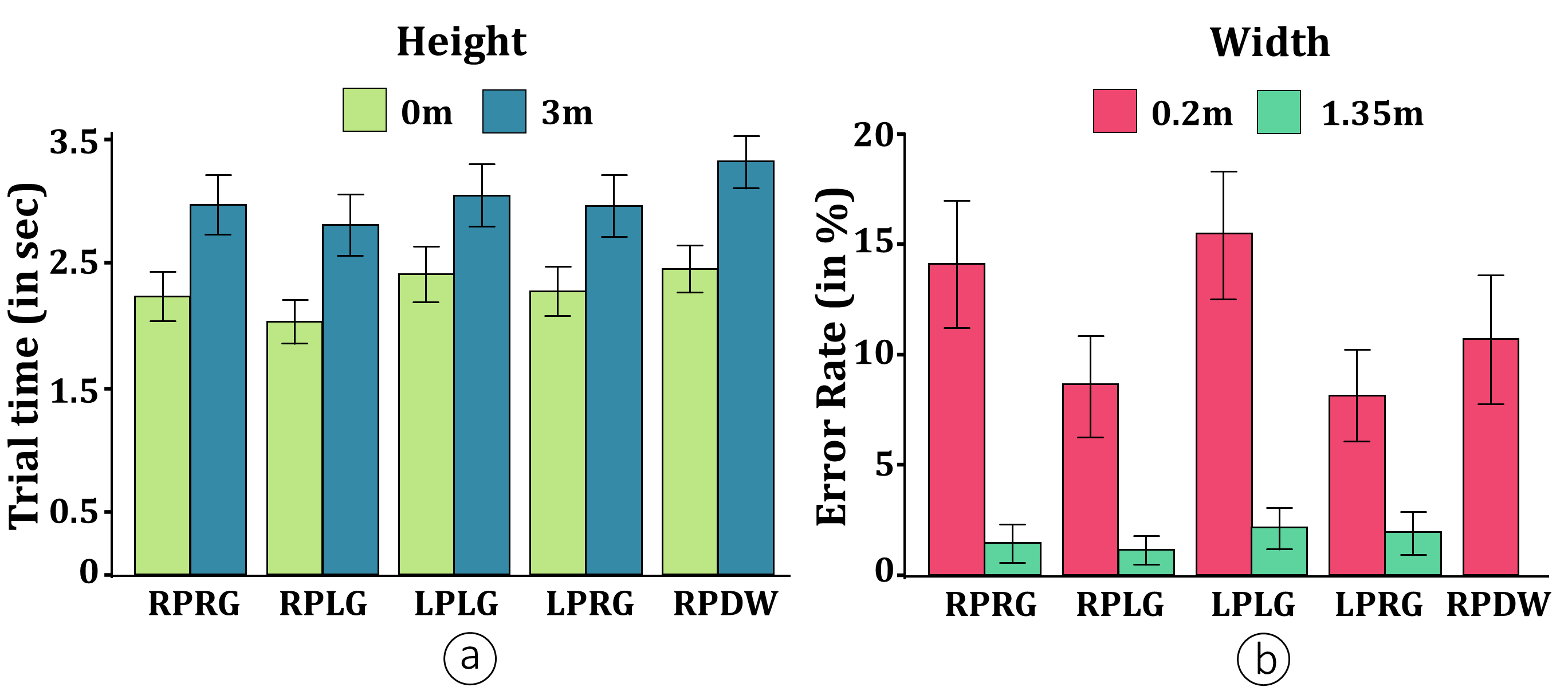}
	\caption{(a) Movement time by \textit{Technique} for each \textit{Height}, (b) Error rate by \textit{Technique} for each \textit{Target Width}. \update{Error bars represent 95\% Confidence Interval.}} 
	\label{fig:tasks} 
\end{figure}

We found significant interactions of \textit{Technique} $\times$ \textit{Height} ($F_{4,72} = 2.97$, $p < 0.05$, $\eta^2 = 0.14$), \textit{Distance} $\times$ \textit{Height} ($F_{1, 18} = 8.09$, $p<0.05$, $\eta^2=0.31$),  \textit{Technique} $\times$ \textit{Width} ($F_{4, 72} = 31.42$, $p < 0.001$, $\eta^2 = 0.64$), \textit{Distance} $\times$ \textit{Width} ($F_{1, 18} = 37.35$, $p < 0.001$, $\eta^2 = 0.67$) and \textit{Height} $\times$ \textit{Width} ($F_{1, 18} = 24.27$, $p < 0.001$, $\eta^2 = 0.57$). No other interaction effect was found.


\subsubsection{Error Rate}
We found significant main effects for the independent variables \textit{Technique} ($F_{4, 76} = 9.89$, $p < 0.001$, $\eta^2$ = 0.34), \textit{Distance} ($F_{1, 19} = 250.62$, $p < 0.001$, $\eta^2$ = 0.93), \textit{Height} ($F_{1, 19} = 88.68$, $p < 0.001$, $\eta^2$ = 0.82), and \textit{Width} ($F_{1, 19}  = 196.98$, $p < 0.001$, $\eta^2$ = 0.91). The mean error rate is 8\% (95\% confidence interval, CI: [6\%, 9\%]) for RPRG, 5\% (CI: [4\%, 6\%]) for RPLG, 9\% (CI: [7\%, 10\%]) for LPLG, 5\% (CI: [4\%, 6\%]) for LPRG and 5\% (CI: [4\%, 7\%]) for RPDW.
Post-hoc pairwise comparisons reveal that error rates with RPLG, LPRG, and RPDW are significantly more accurate (all $p<0.001$) than the uni-manual gesture-based techniques (i.e., LPLG and RPRG).

Naturally, the higher target \textit{Width} (1.35m (M = 1\%, [1\%, 2\%]) leads to significantly accurate selection ($p<0.001$) than lower \textit{Width} (0.2m: M = 11\%, CI: [10\%, 13\%]).  
The lower target \textit{Distance} (3m, M = 1\%, [1\%, 2\%]) leads to significantly more accurate selection ($p<0.001$) than higher target \textit{Distance} (9m, M = 11\%, CI: [10\%, 13\%).
Similarly, lower target \textit{Height} (0m, M = 4\%, CI:[4\%, 5\%]) leads to significantly accurate selection ($p<0.001$) than higher target \textit{Height} (3m, M = 9\%, CI:[7\%, 10\%])

We found significant interactions of \textit{Distance} $\times$ \textit{Height} ($F_{1, 19} = 70.6$, $p<0.001$, $\eta^2=0.79$), \textit{Technique} $\times$ \textit{Width} ($F_{4, 76} = 6.78$, $p < 0.001$, $\eta^2 = 0.26$), \textit{Distance} $\times$ \textit{Width} ($F_{1, 19} = 222.02$, $p < 0.001$, $\eta^2 = 0.92$) and \textit{Height} $\times$ \textit{Width} ($F_{1, 19} = 27.27$, $p < 0.001$, $\eta^2 = 0.59$) on error rate. No other interaction effect was found.

\subsection{Preference Scores}
We collected users' feedback on the seven Nasa TLX criteria along with their preference and fatigue for \textit{Technique} and \textit{Posture}.

  \begin{figure}[h!]
    \centering
    \includegraphics[width=1.0\columnwidth]{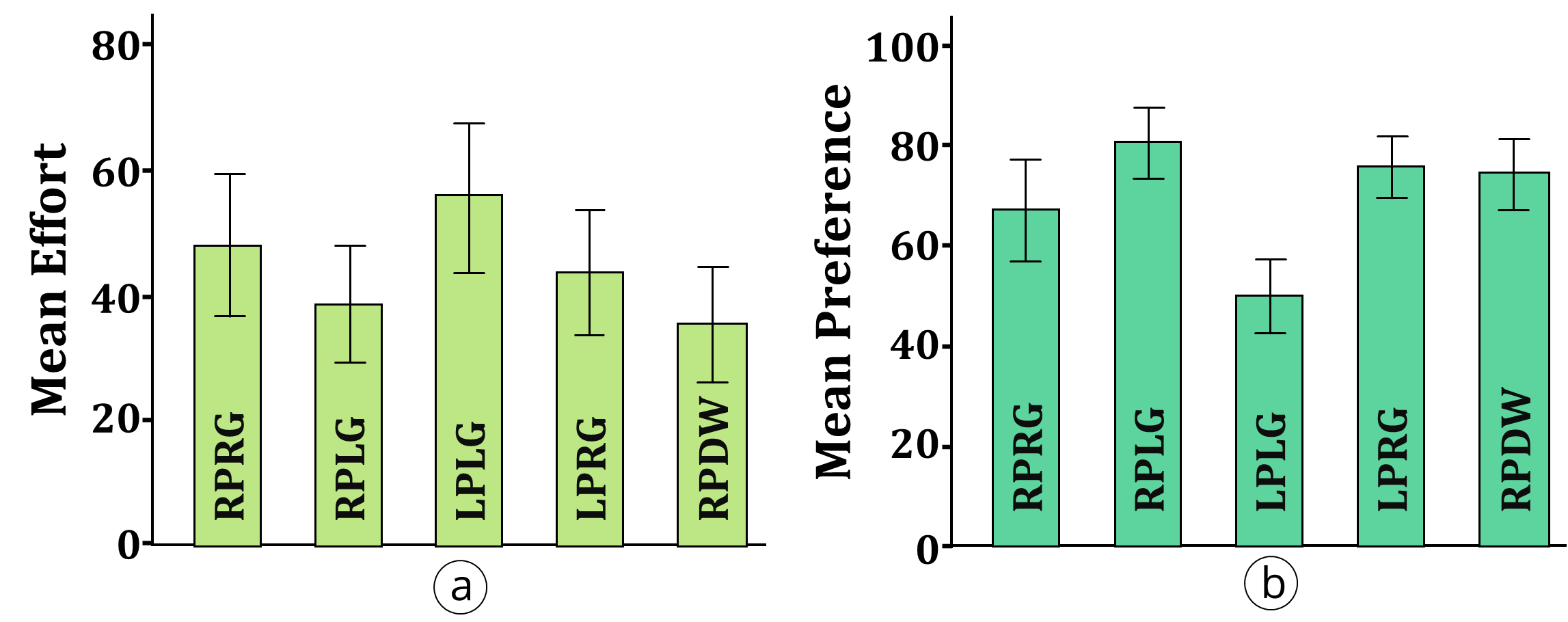}
	\caption{Subjective feedback on each \textit{Technique} (out of 100) scores for (a) Mean Effort (the lower the value, the less the effort), (b) Mean Preference (the higher the value, the more preferred the technique). Error bars represent 95\% Confidence Interval} 
	\label{fig:results_preference} 
\end{figure} 

Friedman tests show that participants there was no subjective difference between the two \textit{Postures}: Sitting and Standing for any of the nine subjective feedback criteria.

For \textit{Technique}, Friedman tests reveal significant differences for seven out of eight subjective criteria [Preference ($\chi^2 (4,N=20)=28.864, p<0.001$), Fatigue ($\chi^2 (4,N=20)=28.506, p<0.001$), Mental Demand ($\chi^2 (4,N=20)=19.78, p<0.001$), Physical Demand ($\chi^2 (4,N=20)=9.52, p<0.05$), Effort ($\chi^2 (4,N=20)=39.13, p<0.001$), Frustration ($\chi^2 (4,N=20)=36.80, p<0.001$), and Overall task load ($\chi^2 (4,N=20)=29.80, p<0.001$)].
However, \textit{Technique} does not have a significant effect on  Performance ($\chi^2 (4,N=20)=23.48, p=0.76$) and Temporal Demand ($\chi^2 (4,N=20)=3.63, p=0.46$).

Post-hoc pairwise comparisons (Bonferroni: $\alpha$-levels from 0.05 to 0.005) reveal that:
LPLG has less Preference, more Fatigue, more mentally demand, more Effort,  more Frustration, and more Overall Demand than RPLG, LPRG, and RPDW. 
 Furthermore, RPRG also requires more Effort than RPDW.

\begin{table*}[]
\caption{Comparison of Fitts' law models for pointing on different conditions.
}

\label{tab:fittsComparison}
\begin{center}
 \begin{adjustbox}{width=0.7\textwidth}
\begin{tabular}{|c|c|cc|c|c|c|c|c|}
\hline
Model    & Condition & \multicolumn{2}{c|}{F-test}                     & R2 & Adj R2 & AIC & BIC & Equation (-b1/b2) \\ \cline{3-4}
                          &                            & \multicolumn{1}{c|}{F-stat} & p-val             &                     &                         &                      &                      &                                    \\ \hline
Standard & RPRG                       & \multicolumn{1}{c|}{60.32}  & p\textless{}0.001 & 0.91                & 0.89                    & 10.49                & 10.65                & MT=0.83ID-0.48                     \\ \cline{2-9} 
                          & LPLG                       & \multicolumn{1}{c|}{80.42}  & p\textless{}0.001 & 0.93                & 0.92                    & 9.94                 & 10.10                & MT= 0.92ID-0.68                    \\ \cline{2-9} 
                          & RPLG                       & \multicolumn{1}{c|}{36.78}  & p\textless{}0.001 & 0.86                & 0.84                    & 14.07                & 14.23                & MT = 0.81ID-0.57                   \\ \cline{2-9} 
                          & LPRG                       & \multicolumn{1}{c|}{42.22}  & p\textless{}0.001 & 0.88                & 0.86                    & 13.96                & 14.12                & MT=0.86*ID-0.55                    \\ \cline{2-9} 
                          & RPDW                       & \multicolumn{1}{c|}{29.67}  & p\textless{}0.01  & 0.83                & 0.80                    & 14.00                & 14.16                & MT=0.72*ID+0.21                    \\ \cline{2-9} 
                          & All Sit                    & \multicolumn{1}{c|}{51.67}  & p\textless{}0.001 & 0.90                & 0.88                    & 11.14                & 11.30                & MT=0.80*ID-0.34                    \\ \cline{2-9} 
                          & All Stand                  & \multicolumn{1}{c|}{46.30}  & p\textless{}0.001 & 0.89                & 0.87                    & 13.20                & 13.36                & MT=0.86*ID-0.49                    \\ \cline{2-9} 
                          & All                        & \multicolumn{1}{c|}{48.92}  & p\textless{}0.001 & 0.89                & 0.87                    & 12.18                & 12.34                & MT=0.83*ID-0.41                    \\ \hline
Two-part & RPRG                       & \multicolumn{1}{c|}{27.98}  & p\textless{}0.01  & 0.92               & 0.89                   & 11.71                & 11.94                & MT=0.99*A+0.51*B-1.43              \\ \cline{2-9} 
                          & LPLG                       & \multicolumn{1}{c|}{37.42}  & p\textless{}0.001 & 0.94               & 0.91                   & 11.11                & 11.35                & MT=1.08*A+0.50*B-1.63              \\ \cline{2-9} 
                          & RPLG                       & \multicolumn{1}{c|}{18.52}  & p\textless{}0.01  & 0.88               & 0.83                   & 14.75                & 14.99                & MT=1.06*A+0.81*B-2.09              \\ \cline{2-9} 
                          & LPRG                       & \multicolumn{1}{c|}{20.44}  & p\textless{}0.01  & 0.89               & 0.85                   & 14.90                & 15.14                & MT=1.09*A+0.73*B-1.92              \\ \cline{2-9} 
                          & RPDW                       & \multicolumn{1}{c|}{21.95}  & p\textless{}0.01  & 0.90               & 0.86                   & 12.01                & 12.25                & MT=1.13*A+1.29*B-2.23              \\ \cline{2-9} 
                          & All Sit                    & \multicolumn{1}{c|}{26.47}  & p\textless{}0.01  & 0.91               & 0.88                   & 11.64                & 11.88                & MT=1.02*A+0.71*B-1.68              \\ \cline{2-9} 
                          & All Stand                  & \multicolumn{1}{c|}{23.95}  & p\textless{}0.01  & 0.91             & 0.87                   & 13.65                & 13.89                & MT=1.12*A+0.82*B-2.05              \\ \cline{2-9} 
                          & All                        & \multicolumn{1}{c|}{25.20}  & p\textless{}0.01  & 0.91               & 0.87                   & 12.65                & 12.89                & MT=1.07*A+0.77*B-1.86              \\ \hline
Vergence & RPRG                       & \multicolumn{1}{c|}{25.50}  & p\textless{}0.01  & 0.91             & 0.88                  & 12.38                & 12.62                & MT=0.81*A+0.06*B-0.56              \\ \cline{2-9} 
                          & LPLG                       & \multicolumn{1}{c|}{34.61}  & p\textless{}0.01  & 0.93               & 0.91                   & 11.70                & 11.94                & MT=0.90*A+0.08*B-0.81              \\ \cline{2-9} 
                          & RPLG                       & \multicolumn{1}{c|}{16.56}  & p\textless{}0.01  & 0.87               & 0.82                   & 15.53                & 15.77                & MT=0.77A+0.16B - 0.80            \\ \cline{2-9} 
                          & LPRG                       & \multicolumn{1}{c|}{18.62}  & p\textless{}0.01  & 0.88               & 0.83                   & 15.56                & 15.80                & MT=0.83A+0.14B - 0.76            \\ \cline{2-9} 
                          & RPDW                       & \multicolumn{1}{c|}{15.82}  & p\textless{}0.01  & 0.86               & 0.81                   & 14.32                & 14.56                & MT=0.65A+0.28B - 0.19            \\ \cline{2-9} 
                          & All Sit                    & \multicolumn{1}{c|}{23.04}  & p\textless{}0.01  & 0.90               & 0.86                   & 12.65                & 12.89                & MT=0.77A+0.13B - 0.52            \\ \cline{2-9} 
                          & All Stand                  & \multicolumn{1}{c|}{20.94}  & p\textless{}0.01  & 0.89                & 0.85                    & 14.62                & 14.86                & MT=0.82A+0.16B - 0.73            \\ \cline{2-9} 
                          & All                        & \multicolumn{1}{c|}{21.98}  & p\textless{}0.01  & 0.90                & 0.86                    & 13.64                & 13.88                & MT=0.79A+0.14B - 0.63            \\ \hline
Proposed & RPRG                       & \multicolumn{1}{c|}{45.88}  & p\textless{}0.001 & 0.95                & 0.93                    & 8.012                & 8.250                & MT=1.20A+2.94B - 2.48            \\ \cline{2-9} 
                          & LPLG                       & \multicolumn{1}{c|}{69.36}  & p\textless{}0.001 & 0.97                & 0.95                    & 6.410                & 6.648                & MT=1.32A+3.06B - 2.78            \\ \cline{2-9} 
                          & RPLG                       & \multicolumn{1}{c|}{22.51}  & p\textless{}0.01  & 0.90                & 0.86                    & 13.36                & 13.60                & MT=1.19A+3.01B - 2.62            \\ \cline{2-9} 
                          & LPRG                       & \multicolumn{1}{c|}{30.97}  & p\textless{}0.01  & 0.93                & 0.89                    & 11.88                & 12.12                & MT=1.31A+3.53B - 2.96            \\ \cline{2-9} 
                          & RPDW                       & \multicolumn{1}{c|}{15.92}  & p\textless{}0.01  & 0.86                & 0.81                    & 14.28                & 14.52                & MT=1.04*A+2.46*B-1.46              \\ \cline{2-9} 
                          & All Sit                    & \multicolumn{1}{c|}{34.31}  & p\textless{}0.01  & 0.93                & 0.91                    & 9.73                 & 9.97                 & MT=1.15*A+2.76*B-2.22              \\ \cline{2-9} 
                          & All Stand                  & \multicolumn{1}{c|}{32.26}  & p\textless{}0.01  & 0.93                & 0.90                    & 11.47                & 11.70                & MT=1.27*A+3.25*B-2.71              \\ \cline{2-9} 
                          & All                        & \multicolumn{1}{c|}{33.35}  & p\textless{}0.01  & 0.93                & 0.90                    & 10.59                & 10.83                & MT=1.21*A+3.00*B-2.46              \\ \hline
\end{tabular}
\end{adjustbox}
\end{center}
\end{table*}

\subsection{Throughput Analysis}
Throughput (TP) has been recognized as an important metric for quantifying the performance of input systems \cite{fitts1954information, iso20009241, zhai2004speed, mackenzie1992fitts}. Within the context of Fitts' law, both movement time and the index of difficulty are influenced by changes in width and amplitude. Notably, throughput considers both speed (as represented by movement time) and accuracy (as indicated by endpoint deviation). Consequently, for a specific interaction or input device, throughput remains consistent. 
To mitigate potential biases in subjective speed-accuracy trade-offs \cite{zhai2004speed} during pointing tasks, we use Crossman’s correction \cite{crossman1957speed} to determine the effective Index of Difficulty, denoted as \(ID_e\). The formula for \(ID_e\) is given by \(ID_e =\log _2(\frac{A_e}{W_e}+1)\). In this equation, \(W_e\) represents the effective width, which is derived from the standard deviation of endpoints: \(W_e = 4.133\times SD\). Additionally, \(A_e\) signifies the average amplitude between the initial and final positions of the pointer for a specific amplitude condition.
In line with recommendations by Wobbrock et al. \cite{wobbrock2011effects}, we use the means-of-means methodology \cite{soukoreff2004towards, iso20009241} for our throughput calculations. For each display curvature, throughput (TP) is computed using the formula \(TP=\frac{1}{N}\sum_{i=1}^N\left(\frac{ID_{e_i}}{M T_i}\right)\), where \(N= 8\) represents the product of the cardinalities of amplitude and width conditions: \(card(A_e) \times card(W_e)\).

From the throughput analysis we found that the mean throughput ($TP$) is 1.24 bits/s (CI: [0.97, 1.51]) for RPRG, 1.47 bits/s (CI: [1.12, 1.81]) for RPLG, 1.18 bits/s (CI: [0.93, 1.44]) for LPLG, 1.30 bits/s (CI: [1.05, 1.55]) for LPRG and 1.20 bits/s (CI: [0.97, 1.42]) for RPDW. This means that based on the performance analysis, RPLG has the best performance.   

\subsection{Fitts' Law Model comparison}
We performed a comparative analysis of the four models: standard, two-part, vergence, and the proposed model for teleportation performance. We grouped our data into different conditions: 1) data grouped by each technique, 2) data grouped by \textit{posture}, and 3) data for all conditions together.
Table \ref{tab:fittsComparison} shows the performance metrics (i.e., $R^2$, Adj $R^2$, F-statistics score, AIC \cite{akaike1974new, batmaz2020effect}, BIC \cite{schwarz1978estimating} values) for each of the Fitts' law models.     
Besides standard fit measured via  Adj $R^2$ \remove{(Adjusted $R^2$ value)}, \textbf{F-tests} are used to compare if added coefficients (i.e., extra degree-of-freedom in the formula) improve the model prediction. 
\remove{\textbf{The Akaike Information Criterion (AIC)}} \update{\textbf{AIC}} \cite{akaike1974new} and the \remove{\textbf{Bayesian Information Criterion (BIC)}} \update{\textbf{BIC}} \cite{schwarz1978estimating} are both metrics to compare between several regression models which consider the fitness of the model while penalizing models for added coefficients. The lower the score for AIC and BIC, the better the model. 
Considering the overall linear and angular measurement data for each model, we compute $\Delta AIC_i$ = AIC of a $i^{th}$ model minus the lowest AIC ($AIC_{min}$) and $\Delta BIC$= Difference between BIC values of two models (see Table \ref{tab:fittsComparison}). 
To evaluate if there are any significant differences between the models, we use  Burnham and Anderson \cite{burnham2004multimodel} criterion ($\Delta AIC_i<2$ = substantial evidence, $2<\Delta AIC_i<4$ = strong evidence, $4<\Delta AIC_i<7$ = less evidence, $\Delta AIC_i>10$ = no evidence) for AIC values and Raftery \cite{raftery1995bayesian} criterion  ($\Delta BIC>10$ = very strong evidence, $6<\Delta BIC<10$ = strong evidence, $2<\Delta BIC<6$ = positive evidence, $0<\Delta BIC<2$ = no evidence) for BIC values.

\subsubsection {Model Comparison Based on Posture}


For both conditions ``All Sit'' and ``All Stand'', the \textbf{Proposed Model} is the best model based on Adj $R^2$, AIC, and BIC values. Based on  Burnham and Anderson \cite{burnham2004multimodel} criterion and Raftery \cite{raftery1995bayesian} criterion, there is positive evidence that the proposed model is better than the other three models. 
For the sitting condition of the proposed model, we found Adj $R^2$: 0.91, AIC: 9.73, and BIC: 9.97.
The difference in AIC and BIC is at least 2 unit lower than the other models for the sitting condition (Standard Model Adj $R^2$: 0.88, AIC: 11.14, BIC: 11.30; Two-part Model:
Adj $R^2$: 0.88, AIC: 11.64, BIC: 11.88; Vergence-Accomodation Model:
Adj $R^2$: 0.86, AIC: 12.65, BIC: 12.89).
For the standing condition of the proposed model we found Adj $R^2$: 0.90, AIC: 11.47, and BIC: 11.70.
The difference in AIC and BIC is at least 2 unit lower than the other models for the standing condition (Standard Model Adj $R^2$: 0.87, AIC: 13.20BIC: 13.36; Two-part Model:
Adj $R^2$: 0.87, AIC: 13.65, BIC: 13.89; Vergence-Accomodation Model:
Adj $R^2$: 0.85, AIC: 14.62, BIC: 14.86), signifying positive evidence of better model performance.

\subsubsection {Model Comparison Based on All Conditions}


For the condition ``All'', considering Adj $R^2$, AIC, and BIC values, the \textbf{Proposed Model} is the best model for the condition ``All''. For the "All" condition of the proposed model, we found Adj $R^2$: 0.90, AIC: 10.59, and BIC: 10.83.
The difference in AIC and BIC is at least 2 lower than the other models for the "All" condition (Standard Model Adj $R^2$: 0.87, AIC: 12.18, BIC: 12.34; Two-part Model:
Adj $R^2$: 0.87, AIC: 12.65, BIC: 12.89; Vergence-Accomodation Model:
Adj $R^2$: 0.86, AIC: 13.64, BIC: 13.88) which has positive support for the \textbf{Proposed Model} as the superior model.

\subsubsection{Model Comparison Based on Technique}

Except for the ``RPDW'' technique, the \textbf{Proposed Model} consistently outperforms the other four techniques (``RPRG'', ``LPLG'', ``RPLG'', and ``LPRG'') based on Adj $R^2$, AIC, and BIC values as shown in Table \ref{tab:fittsComparison}.

\section{DISCUSSION}
We address our research questions and provide takeaways from our user study. 

\textbf{RQ1: most effective teleportation technique(s):} Our study results showed that bi-manual \textit{techniques} lead to significantly better performance than uni-manual \textit{techniques} in terms of both movement time and accuracy. Interestingly, among the two bi-manual \textit{techniques}, RPLG consistently outperforms LPRG. We attribute this to the natural alignment of certain tasks, like pinpointing and stability, with the dominant hand. This is in line with the findings of previous evaluations of bi-manual \textit{techniques} for 3D interaction \cite{caputo2017single, lubos20144}, and Walk-through \cite{zhang2017double} in VR Environments. 

RPDW, which is a dwell-based \textit{technique}, was found to be as accurate as bi-manual \textit{techniques}. Moreover, RPDW performed well in accuracy for easier tasks, such as targets located at lower \textit{distances} and \textit{heights}. However, it's crucial to note that RPDW demonstrated the highest movement time among all \textit{techniques}, a difference that became more pronounced for easier tasks. This can be attributed to the dwell time one must wait to \remove{trigger}\update{confirm} a selection. In the case of easier tasks, other \textit{techniques} take significantly less time as the participant is free to perform the confirmation gesture to teleport as soon as the pointer is relocated over the target - eliminating the waiting time associated with the dwell technique.  

\textbf{RQ2 - Impact of posture:} We found that \textit{posture} does not affect performance (i.e., movement time and accuracy) or subjective ratings. Furthermore, \textit{posture} did not have any interaction effect with any of the other factors. This confirms the results of previous studies \cite{coomer2018virtual, zielasko2020can, zielasko2020either} that both sitting and standing can be considered for 3D interaction in VR. Users who preferred sitting cited increased comfort, safety, and reduced fatigue, while those favoring standing reported greater freedom of movement, better control, and a more immersive experience.

\textbf{RQ3 - Fitts' law model:} Interestingly, all four models (Proposed Model, Standard Model, Two-part Model, and Vergence-Accommodation Model)  manage to capture our teleportation tasks relatively accurately (all adj $R^2>0.8$), thus showing that Fitts' law and its variations predict teleportation performance. 
However, we observe that our \textbf{proposed model} outperforms the other models in seven out of eight comparisons based on AIC, BIC, and Adj $R^2$. Therefore, we suggest researchers use the \textbf{proposed model} for pointing tasks that involve distant and elevated targets, which shows significantly better performance (substantial evidence based on AIC values and strong to positive evidence based on BIC values) compared to the other models.

\section{Design Implications}

We interpret key insights from our results and discuss the design implications for teleportation in VR.

\textbf{Using Bi-manual Techniques}: Considering the superior performance of bi-manual \textit{techniques} over uni-manual ones, VR designers are encouraged to incorporate bi-manual \textit{techniques} for teleportation in VR applications. Particularly, the RPLG \textit{technique} consistently outperforms other methods, indicating that aligning tasks with the dominant hand can significantly enhance user experience and efficiency. Therefore, we recommend designers to prioritize the use of this technique for VR teleportation applications.

\textbf{Adoption of the Proposed Model for VR}: We introduced a novel model that considers key aspects of VR, including the Vergence-accommodation conflict and targets positioned at different \textit{distances} and \textit{heights}. Our model has exhibited superior performance compared to existing models in VR teleportation, as evidenced by higher scores in Adj $R^2$, AIC, and BIC. We encourage researchers and designers to adopt our proposed model when evaluating VR teleportation. It provides a more comprehensive and accurate representation of user behavior in VR environments, leading to more effective design solutions.
    
 \textbf{Reconsider Dwell Techniques}: While the RPDW \textit{technique}, a dwell-based method, demonstrated high accuracy, especially for simpler tasks, it exhibited significantly higher movement time. Designers should exercise caution when using dwell \textit{techniques}, particularly for applications where speed is crucial. The inherent delay in dwell-based techniques, caused by the necessary hold time, can significantly affect user efficiency.
    
  \textbf{Accommodate Both Sitting and Standing Postures}: The study found no significant difference in performance between sitting and standing \textit{postures}. VR applications should be adaptable to both \textit{postures}, allowing users to choose based on their comfort and the type of experience they seek. This flexibility can cater to a broader audience and enhance user satisfaction.

%% file: 7_guidelines.tex
\section{Limitations and Future work}
We highlight limitations observed during the studies and discuss potential future works.

Our evaluation of bi-manual input and \textit{posture} for VR Teleportation has some limitations. \remove{Our current implementation of bi-manual techniques does not take into account the differences in user's physical attributes such as their height and arms length, which might affect task performance. Further studies can be conducted to explore the personalization of hands-only techniques to mitigate these differences in users' physical attributes.}
We used the `closing the index finger' gestures as it was used in a prior work \cite{chowdhury2022wriarm} for uni-manual gesture-based teleportation. Subsequent research could explore various methods of integrating bi-manual input for selection, including diverse gesture sets involving simultaneous inputs from both hands. 

Although the investigated techniques exhibited promise for teleportation, the error rates were relatively high, with mean error rates ranging from 5\% to 9\% for hands-only teleportation. We attribute this high error rate to our challenging task conditions, such as very small targets, high, and distant targets that pushed the boundaries of the reachable area. Smaller target widths were primarily responsible for the higher error rates, as the error rates for larger targets were close to 1\% for all techniques, as observed in Figure \ref{fig:tasks}(b). Future research could explore precise selection methods to reduce this error rate. We \remove{explored} \update{covered a broad range of Index of Difficulty while exploring a limited set of target sizes, distances, and elevations for our Fitts' law study, based on previous research and pilot studies. This approach excluded a comprehensive investigation of all potential combinations. Furthermore, we recognize that our sample size (N=20) may not be representative of the broader population. Future studies could expand on this by incorporating a wider range of combinations and a larger participant pool. Also, our experiments were not driven by prior hypotheses; thus, results are exploratory in nature.}

Research in the field of VR locomotion has led to the design of diverse setups and postures, including physically walking, walking in place, and leaning, as discussed in studies such as Wilson et al. (2016) \cite{wilson2016vr}. However, these aspects were beyond the scope of our study. Future research could consider these setups to conduct a more comprehensive exploration of how users' teleportation performance might be influenced by these factors. We also did not include multiple teleportation steps and other pointer types (e.g., straight line pointer). A straight-line pointer has been shown to have advantages for mid-air targets \cite{matviienko2022skyport}; its effectiveness to platform-based destinations can be explored in a future study.

\update{Hand-tracking techniques necessitate users to keep their hands raised and visible to the headset's cameras, which can lead to ``gorilla arm syndrome", a condition caused by prolonged elevation of the arms. Our study did not look at the long-term usage of hands-free techniques on teleportation, which could be a future venue topic to explore. For instance, researchers could investigate a metric to quantify arm fatigue of VR interactions and find the ergonomic feasibility of hand-based teleportation.}

\update{We adopted a simplified task to help participants focus on teleportation without visually searching for targets located off-screen - aligning with the design used in Fitts' law \cite{barrera2019effect, shoemaker2012two}. We used this setup to provide initial insights into the effectiveness of the bi-manual \textit{techniques} and encourage future studies to incorporate more complex navigation tasks, including virtual rotation and visual search.}



%% file: 8_conclusion.tex
\section{Conclusion}
Our study investigated the effects of hand input and posture while performing teleportation in VR. We explored a set of uni-manual and bi-manual techniques for hand-based teleportation in both seated and standing postures. 
With a user study, we found that posture did not have any significant effect on performance; however, we found that both movement time and accuracy were significantly better for bi-manual techniques as compared to uni-manual ones. The dominant hand for pointing and the non-dominant hand for gesture (i.e., RPLG) resulted in the fastest movement time while ensuring high accuracy. Although the dwell technique (i.e., RPDW) has the slowest movement time, it had similar levels of accuracy. In addition, we proposed a new Fitts' law model, which performed better than three popular prior Fitts' law models used in VR. We finally end with guidelines for teleportation in VR.






%% file: main.bbl
\begin{thebibliography}{10}

\bibitem{akaike1974new}
H.~Akaike.
\newblock A new look at the statistical model identification.
\newblock {\em IEEE Transactions on Automatic Control}, 19(6):716--723, 1974. \href{https://doi.org/10.1109/TAC.1974.1100705}
{doi: {{%
10\hspace{.1pt}\discretionary{.}{%
}{.}\hspace{.4pt}1109\discretionary{/}{%
}{/}TAC\hspace{.1pt}\discretionary{.}{%
}{.}\hspace{.4pt}1974\hspace{.1pt}\discretionary{.}{%
}{.}\hspace{.4pt}1100705}}}


\bibitem{barrera2019effect}
M.~D. Barrera~Machuca and W.~Stuerzlinger.
\newblock The effect of stereo display deficiencies on virtual hand pointing.
\newblock In {\em Proc. of the 2019 CHI Conf. on Human Factors in Computing Systems}, p. 1–14. ACM, 2019. \href{https://doi.org/10.1145/3290605.3300437}
{doi: {{%
10\hspace{.1pt}\discretionary{.}{%
}{.}\hspace{.4pt}1145\discretionary{/}{%
}{/}3290605\hspace{.1pt}\discretionary{.}{%
}{.}\hspace{.4pt}3300437}}}


\bibitem{batmaz2020effect}
A.~U. Batmaz and W.~Stuerzlinger.
\newblock Effect of fixed and infinite ray length on distal 3{D} pointing in virtual reality.
\newblock In {\em Extended Abstracts of the 2020 CHI Conf. on Human Factors in Computing Systems}, CHI EA '20, p. 1–10. ACM, USA, 2020. \href{https://doi.org/10.1145/3334480.3382796}
{doi: {{%
10\hspace{.1pt}\discretionary{.}{%
}{.}\hspace{.4pt}1145\discretionary{/}{%
}{/}3334480\hspace{.1pt}\discretionary{.}{%
}{.}\hspace{.4pt}3382796}}}


\bibitem{bhandari2018teleportation}
J.~Bhandari, P.~R. MacNeilage, and E.~Folmer.
\newblock Teleportation without spatial disorientation using optical flow cues.
\newblock In {\em Graphics interface}, pp. 162--167, 2018.

\bibitem{boletsis2019vr}
C.~Boletsis and J.~E. Cedergren.
\newblock {VR} locomotion in the new era of virtual reality: an empirical comparison of prevalent techniques.
\newblock {\em Advances in Human-Computer Interaction}, 2019.

\bibitem{bolte2011jumper}
B.~Bolte, F.~Steinicke, and G.~Bruder.
\newblock The jumper metaphor: an effective navigation technique for immersive display setups.
\newblock In {\em Proc. of Virtual Reality Int. Conf.}, vol.~1, 2011.

\bibitem{bowman20043d}
D.~Bowman, E.~Kruijff, J.~J. LaViola~Jr, and I.~P. Poupyrev.
\newblock {\em 3{D} User interfaces: theory and practice, CourseSmart eTextbook}.
\newblock Addison-Wesley, 2004.

\bibitem{bowman1997travel}
D.~A. Bowman, D.~Koller, and L.~F. Hodges.
\newblock Travel in immersive virtual environments: An evaluation of viewpoint motion control techniques.
\newblock In {\em Proceedings of IEEE 1997 Annual International Symposium on Virtual Reality}, pp. 45--52. IEEE, 1997.

\bibitem{bozgeyikli2016a}
E.~Bozgeyikli, A.~Raij, S.~Katkoori, and R.~Dubey.
\newblock Locomotion in virtual reality for individuals with autism spectrum disorder.
\newblock In {\em Proc. of the 2016 Symposium on Spatial User Interaction}, 2016.

\bibitem{bozgeyikli2016b}
E.~Bozgeyikli, A.~Raij, S.~Katkoori, and R.~Dubey.
\newblock Point \& teleport locomotion technique for virtual reality.
\newblock In {\em Proc. of the 2016 Annual Symposium on Computer-Human Interaction in Play (CHI PLAY '16)}, pp. 205--216. ACM, 2016. \href{https://doi.org/10.1145/2967934.2968105}
{doi: {{%
10\hspace{.1pt}\discretionary{.}{%
}{.}\hspace{.4pt}1145\discretionary{/}{%
}{/}2967934\hspace{.1pt}\discretionary{.}{%
}{.}\hspace{.4pt}2968105}}}


\bibitem{burnham2004multimodel}
K.~P. Burnham and D.~R. Anderson.
\newblock Multimodel inference: Understanding aic and bic in model selection.
\newblock {\em Sociological Methods \& Research}, 33(2):261--304, 2004. \href{https://doi.org/10.1177/0049124104268644}
{doi: {{%
10\hspace{.1pt}\discretionary{.}{%
}{.}\hspace{.4pt}1177\discretionary{/}{%
}{/}0049124104268644}}}


\bibitem{caputo2017single}
F.~M. Caputo, M.~Emporio, A.~Giachetti, et~al.
\newblock Single-handed vs. two handed manipulation in virtual reality: A novel metaphor and experimental comparisons.
\newblock In {\em STAG}, pp. 39--45, 2017.

\bibitem{chastine2013study}
J.~Chastine, N.~Kosoris, and J.~Skelton.
\newblock A study of gesture-based first person control.
\newblock In {\em Proc. of {CGAMES}'2013 {USA}}, pp. 79--86. IEEE, 2013.

\bibitem{cherni2020literature}
H.~Cherni, N.~M{\'e}tayer, and N.~Souliman.
\newblock Literature review of locomotion techniques in virtual reality.
\newblock {\em Int. J. of Virtual Reality}, 2020.

\bibitem{chowdhury2023paws}
S.~Chowdhury, W.~Delamare, P.~Irani, and K.~Hasan.
\newblock Paws: Personalized arm and wrist movements with sensitivity mappings for controller-free locomotion in virtual reality.
\newblock {\em Proc. of the ACM on Human-Computer Interaction}, 7(MHCI):1--21, 2023.

\bibitem{chowdhury2022wriarm}
S.~Chowdhury, A.~K. M.~A. Ullah, N.~B. Pelmore, P.~Irani, and K.~Hasan.
\newblock Wriarm: Leveraging wrist movement to design wrist+arm based teleportation in {VR}.
\newblock In {\em 2022 IEEE Int. Symposium on Mixed and Augmented Reality ({ISM{AR}})}, pp. 317--325, 2022. \href{https://doi.org/10.1109/ISM{AR}55827.2022.00047}
{doi: {{%
10\hspace{.1pt}\discretionary{.}{%
}{.}\hspace{.4pt}1109\discretionary{/}{%
}{/}ISM{AR}55827\hspace{.1pt}\discretionary{.}{%
}{.}\hspace{.4pt}2022\hspace{.1pt}\discretionary{.}{%
}{.}\hspace{.4pt}0}}}


\bibitem{christou2017steering}
C.~G. Christou and P.~Aristidou.
\newblock Steering versus teleport locomotion for head mounted displays.
\newblock In {\em Int. Conf. on augmented reality, virtual reality and computer graphics}, pp. 431--446. Springer, 2017.

\bibitem{clifton2020effects}
J.~Clifton and S.~Palmisano.
\newblock Effects of steering locomotion and teleporting on cybersickness and presence in hmd-based virtual reality.
\newblock {\em Virtual Reality}, 24(3):453--468, 2020.

\bibitem{coomer2018evaluating}
N.~Coomer, S.~Bullard, W.~Clinton, and B.~Williams-Sanders.
\newblock Evaluating the effects of four {VR} locomotion methods: Joystick, arm-cycling, point-tugging, and teleporting.
\newblock In {\em Proc. of the 15th {ACM} Symposium on Applied Perception ({SAP} '18)}, pp. Article 7, 8 pages. ACM, 2018. \href{https://doi.org/10.1145/3225153.3225175}
{doi: {{%
10\hspace{.1pt}\discretionary{.}{%
}{.}\hspace{.4pt}1145\discretionary{/}{%
}{/}3225153\hspace{.1pt}\discretionary{.}{%
}{.}\hspace{.4pt}3225175}}}


\bibitem{coomer2018virtual}
N.~Coomer, J.~Ladd, and B.~Williams.
\newblock Virtual exploration: Seated versus standing.
\newblock In {\em VISIGRAPP (1: GRAPP)}, pp. 264--272, 2018.

\bibitem{crossman1957speed}
E.~Crossman.
\newblock The speed and accuracy of simple hand movements.
\newblock {\em The nature and acquisition of industrial skills}, 1957.

\bibitem{crossman1956measurement}
E.~R. Crossman.
\newblock {\em The measurement of perceptual load in manual operations}.
\newblock PhD thesis, University of Birmingham, 1956.

\bibitem{ferracani2016locomotion}
A.~Ferracani, D.~Pezzatini, J.~Bianchini, G.~Biscini, and A.~Del~Bimbo.
\newblock Locomotion by natural gestures for immersive virtual environments.
\newblock In {\em Proc. of the 1st international workshop on multimedia alternate realities}, pp. 21--24, 2016.

\bibitem{fitts1954information}
P.~M. Fitts.
\newblock The information capacity of the human motor system in controlling the amplitude of movement.
\newblock {\em J. of experimental psychology}, 47(6):381, 1954. \href{https://doi.org/10.1037/h0055392}
{doi: {{%
10\hspace{.1pt}\discretionary{.}{%
}{.}\hspace{.4pt}1037\discretionary{/}{%
}{/}h0055392}}}


\bibitem{folmer2021teleportation}
E.~Folmer, I.~B. Adhanom, and A.~Prithul.
\newblock Teleportation in virtual reality; a mini-review.
\newblock {\em Frontiers in Virtual Reality}, p. 138, 2021.

\bibitem{franzluebbers2023versatile}
A.~Franzluebbers and K.~Johnsen.
\newblock Versatile mixed-method locomotion under free-hand and controller-based virtual reality interfaces.
\newblock In {\em Proceedings of the 29th ACM Symposium on Virtual Reality Software and Technology}, pp. 1--10, 2023.

\bibitem{frommel2017effects}
J.~Frommel, S.~Sonntag, and M.~Weber.
\newblock Effects of controller-based locomotion on player experience in a virtual reality exploration game.
\newblock In {\em Proc. of the 12th Int. Conf. on the Foundations of Digital Games (FDG '17)}, pp. Article 30, 6 pages. ACM, 2017. \href{https://doi.org/10.1145/3102071.3102082}
{doi: {{%
10\hspace{.1pt}\discretionary{.}{%
}{.}\hspace{.4pt}1145\discretionary{/}{%
}{/}3102071\hspace{.1pt}\discretionary{.}{%
}{.}\hspace{.4pt}3102082}}}


\bibitem{funk2019assessing}
M.~Funk, F.~M{\"u}ller, M.~Fendrich, M.~Shene, M.~Kolvenbach, N.~Dobbertin, S.~G{\"u}nther, and M.~M{\"u}hlh{\"a}user.
\newblock Assessing the accuracy of point \& teleport locomotion with orientation indication for virtual reality using curved trajectories.
\newblock In {\em Proc. of the 2019 CHI Conf. on Human Factors in Computing Systems (CHI '19)}, pp. 1--12. ACM, 2019. \href{https://doi.org/10.1145/3290605.3300377}
{doi: {{%
10\hspace{.1pt}\discretionary{.}{%
}{.}\hspace{.4pt}1145\discretionary{/}{%
}{/}3290605\hspace{.1pt}\discretionary{.}{%
}{.}\hspace{.4pt}3300377}}}


\bibitem{griffin2018evaluation}
N.~N. Griffin, J.~Liu, and E.~Folmer.
\newblock Evaluation of handsbusy vs handsfree virtual locomotion.
\newblock In {\em Proceedings of the 2018 Annual Symposium on Computer-Human Interaction in Play}, pp. 211--219, 2018.

\bibitem{grossman2004pointing}
T.~Grossman and R.~Balakrishnan.
\newblock Pointing at trivariate targets in 3{D} environments.
\newblock In {\em Proc. of the SIGCHI Conf. on Human Factors in Computing Systems}, CHI '04, p. 447–454. ACM, USA, 2004. \href{https://doi.org/10.1145/985692.985749}
{doi: {{%
10\hspace{.1pt}\discretionary{.}{%
}{.}\hspace{.4pt}1145\discretionary{/}{%
}{/}985692\hspace{.1pt}\discretionary{.}{%
}{.}\hspace{.4pt}985749}}}


\bibitem{hoffman2008vergence}
D.~M. Hoffman, A.~R. Girshick, K.~Akeley, and M.~S. Banks.
\newblock {Vergence–accommodation conflicts hinder visual performance and cause visual fatigue}.
\newblock {\em J. of Vision}, 8(3):33--33, 03 2008. \href{https://doi.org/10.1167/8.3.33}
{doi: {{%
10\hspace{.1pt}\discretionary{.}{%
}{.}\hspace{.4pt}1167\discretionary{/}{%
}{/}8\hspace{.1pt}\discretionary{.}{%
}{.}\hspace{.4pt}3\hspace{.1pt}\discretionary{.}{%
}{.}\hspace{.4pt}33}}}


\bibitem{huang2019design}
R.~Huang, C.~Harris-Adamson, D.~Odell, and D.~Rempel.
\newblock Design of finger gestures for locomotion in virtual reality.
\newblock {\em Virtual Reality \& Intelligent Hardware}, 1(1):1--9, 2019.

\bibitem{iso20009241}
I.~ISO.
\newblock 9241-9 ergonomic requirements for office work with visual display terminals (vdts)-part 9: Requirements for non-keyboard input devices (fdis-final draft int. standard), 2000.
\newblock {\em Int. Organization for Standardization}, 2000.

\bibitem{janzen2016modeling}
I.~Janzen, V.~K. Rajendran, and K.~S. Booth.
\newblock Modeling the impact of depth on pointing performance.
\newblock In {\em Proc. of the 2016 CHI Conf. on Human Factors in Computing Systems}, CHI '16, p. 188–199. ACM, USA, 2016. \href{https://doi.org/10.1145/2858036.2858244}
{doi: {{%
10\hspace{.1pt}\discretionary{.}{%
}{.}\hspace{.4pt}1145\discretionary{/}{%
}{/}2858036\hspace{.1pt}\discretionary{.}{%
}{.}\hspace{.4pt}2858244}}}


\bibitem{khundam2015first}
C.~Khundam.
\newblock First person movement control with palm normal and hand gesture interaction in virtual reality.
\newblock In {\em 2015 12th Int. Joint Conf. on Computer Science and Software Engineering (JCSSE)}, pp. 325--330. IEEE, 2015.

\bibitem{kopper2010human}
R.~Kopper, D.~A. Bowman, M.~G. Silva, and R.~P. McMahan.
\newblock A human motor behavior model for distal pointing tasks.
\newblock {\em Int. J. of Human-Computer Studies}, 68(10):603--615, 2010. \href{https://doi.org/10.1016/j.ijhcs.2010.05.001}
{doi: {{%
10\hspace{.1pt}\discretionary{.}{%
}{.}\hspace{.4pt}1016\discretionary{/}{%
}{/}j\hspace{.1pt}\discretionary{.}{%
}{.}\hspace{.4pt}ijhcs\hspace{.1pt}\discretionary{.}{%
}{.}\hspace{.4pt}2010\hspace{.1pt}\discretionary{.}{%
}{.}\hspace{.4pt}05\hspace{.1pt}\discretionary{.}{%
}{.}\hspace{.4pt}001}}}


\bibitem{lai2021cognitive}
C.~Lai, X.~Hu, A.~A. Aiyaz, A.~Segismundo, A.~Phadke, and R.~P. McMahan.
\newblock The cognitive loads and usability of target-based and steering-based travel techniques.
\newblock {\em IEEE Transactions on Visualization and Computer Graphics}, 27(11):4289--4299, 2021.

\bibitem{langbehn2018evaluation}
E.~Langbehn, P.~Lubos, and F.~Steinicke.
\newblock Evaluation of locomotion techniques for room-scale {VR}: Joystick, teleportation, and redirected walking.
\newblock In {\em Proc. of the Virtual Reality Int. Conf.-Laval Virtual}, pp. 1--9, 2018.

\bibitem{leyrer2011influence}
M.~Leyrer, S.~A. Linkenauger, H.~H. B{\"u}lthoff, U.~Kloos, and B.~Mohler.
\newblock The influence of eye height and avatars on egocentric distance estimates in immersive virtual environments.
\newblock In {\em Proc. of the ACM SIGGRAPH symposium on applied perception in graphics and visualization}, pp. 67--74, 2011.

\bibitem{leyrer2015eye}
M.~Leyrer, S.~A. Linkenauger, H.~H. B{\"u}lthoff, and B.~J. Mohler.
\newblock Eye height manipulations: A possible solution to reduce underestimation of egocentric distances in head-mounted displays.
\newblock {\em ACM Transactions on Applied Perception (TAP)}, 12(1):1--23, 2015.

\bibitem{lubos20144}
P.~Lubos, G.~Bruder, and F.~Steinicke.
\newblock Are 4 hands better than 2? bimanual interaction for quadmanual user interfaces.
\newblock In {\em Proc. of the 2nd ACM symposium on Spatial user interaction}, pp. 123--126, 2014.

\bibitem{mackenzie1992fitts}
I.~S. MacKenzie.
\newblock Fitts' law as a research and design tool in human-computer interaction.
\newblock {\em Hum.-Comput. Interact.}, 7(1):91–139, mar 1992. \href{https://doi.org/10.1207/s15327051hci0701_3}
{doi: {{%
10\hspace{.1pt}\discretionary{.}{%
}{.}\hspace{.4pt}1207\discretionary{/}{%
}{/}s15327051hci0701\_3}}}


\bibitem{marchal2011joyman}
M.~Marchal, J.~Pettr{\'e}, and A.~L{\'e}cuyer.
\newblock Joyman: A human-scale joystick for navigating in virtual worlds.
\newblock In {\em 2011 IEEE Symposium on 3{D} User Interfaces ( 3{D}UI)}, pp. 19--26. IEEE, 2011.

\bibitem{matviienko2022skyport}
A.~Matviienko, F.~M{\"u}ller, M.~Schmitz, M.~Fendrich, and M.~M{\"u}hlh{\"a}user.
\newblock Skyport: Investigating 3{D} teleportation methods in virtual environments.
\newblock In {\em CHI Conf. on Human Factors in Computing Systems (CHI '22)}, pp. Article 516, 11 pages. ACM, 2022. \href{https://doi.org/10.1145/3491102.3501983}
{doi: {{%
10\hspace{.1pt}\discretionary{.}{%
}{.}\hspace{.4pt}1145\discretionary{/}{%
}{/}3491102\hspace{.1pt}\discretionary{.}{%
}{.}\hspace{.4pt}3501983}}}


\bibitem{mine1995virtual}
M.~R. Mine.
\newblock Virtual environment interaction techniques.
\newblock Technical report, UNC Chapel Hill CS Dept, 1995.

\bibitem{namnakani2023comparing}
O.~Namnakani, Y.~Abdrabou, J.~Grizou, A.~Esteves, and M.~Khamis.
\newblock Comparing dwell time, pursuits and gaze gestures for gaze interaction on handheld mobile devices.
\newblock In {\em Proc. of the 2023 CHI Conf. on Human Factors in Computing Systems}, pp. 1--17, 2023.

\bibitem{nancel2011mid}
M.~Nancel, J.~Wagner, E.~Pietriga, O.~Chapuis, and W.~Mackay.
\newblock Mid-air pan-and-zoom on wall-sized displays.
\newblock In {\em Proceedings of the SIGCHI Conference on Human Factors in Computing Systems}, pp. 177--186, 2011.

\bibitem{pai2017armswing}
Y.~S. Pai and K.~Kunze.
\newblock Armswing: Using arm swings for accessible and immersive navigation in {AR}/vr spaces.
\newblock In {\em Proc. of the 16th Int. Conf. on Mobile and Ubiquitous Multimedia}, pp. 189--198, 2017.

\bibitem{prithul2022evaluation}
A.~Prithul, J.~Bhandari, W.~Spurgeon, and E.~Folmer.
\newblock Evaluation of hands-free teleportation in {VR}.
\newblock In {\em Proc. of the 2022 ACM Symposium on Spatial User Interaction}, pp. 1--6, 2022.

\bibitem{raftery1995bayesian}
A.~E. Raftery.
\newblock Bayesian model selection in social research.
\newblock {\em Sociological methodology}, pp. 111--163, 1995.

\bibitem{rebsdorf2023blink}
M.~R. Rebsdorf, T.~Khumsan, J.~Valvik, N.~C. Nilsson, and A.~Adjorlu.
\newblock Blink don't wink: Exploring blinks as input for {VR} games.
\newblock In {\em Proceedings of the 2023 ACM Symposium on Spatial User Interaction}, pp. 1--8, 2023.

\bibitem{sarupuri2020testbed}
B.~Sarupuri, S.~Jung, S.~Hoermann, M.~C. Whitton, and R.~W. Lindeman.
\newblock Testbed evaluation of multi-travel mode in virtual reality.
\newblock 2020.

\bibitem{schafer2021controlling}
A.~Sch{\"a}fer, G.~Reis, and D.~Stricker.
\newblock Controlling teleportation-based locomotion in virtual reality with hand gestures: a comparative evaluation of two-handed and one-handed techniques.
\newblock {\em Electronics}, 10(6):715, 2021.

\bibitem{schafer2022controlling}
A.~Sch{\"a}fer, G.~Reis, and D.~Stricker.
\newblock Controlling continuous locomotion in virtual reality with bare hands using hand gestures.
\newblock In {\em Int. Conf. on Virtual Reality and Mixed Reality}, pp. 191--205. Springer, 2022.

\bibitem{schwarz1978estimating}
G.~Schwarz.
\newblock Estimating the dimension of a model.
\newblock {\em The Annals of Statistics}, 6(2):461--464, 1978.

\bibitem{shoemaker2012two}
G.~Shoemaker, T.~Tsukitani, Y.~Kitamura, and K.~S. Booth.
\newblock Two-part models capture the impact of gain on pointing performance.
\newblock {\em ACM Trans. Comput.-Hum. Interact.}, 19(4), dec 2012. \href{https://doi.org/10.1145/2395131.2395135}
{doi: {{%
10\hspace{.1pt}\discretionary{.}{%
}{.}\hspace{.4pt}1145\discretionary{/}{%
}{/}2395131\hspace{.1pt}\discretionary{.}{%
}{.}\hspace{.4pt}2395135}}}


\bibitem{soukoreff2004towards}
R.~W. Soukoreff and I.~S. MacKenzie.
\newblock Towards a standard for pointing device evaluation, perspectives on 27 years of fitts’ law research in hci.
\newblock {\em Int. J. of Human-Computer Studies}, 61(6):751--789, 2004.
\newblock Fitts' law 50 years later: applications and contributions from human-computer interaction. \href{https://doi.org/10.1016/j.ijhcs.2004.09.001}
{doi: {{%
10\hspace{.1pt}\discretionary{.}{%
}{.}\hspace{.4pt}1016\discretionary{/}{%
}{/}j\hspace{.1pt}\discretionary{.}{%
}{.}\hspace{.4pt}ijhcs\hspace{.1pt}\discretionary{.}{%
}{.}\hspace{.4pt}2004\hspace{.1pt}\discretionary{.}{%
}{.}\hspace{.4pt}09\hspace{.1pt}\discretionary{.}{%
}{.}\hspace{.4pt}001}}}


\bibitem{teather2014visual}
R.~J. Teather and W.~Stuerzlinger.
\newblock Visual aids in 3{D} point selection experiments.
\newblock In {\em Proc. of the 2nd ACM symposium on Spatial user interaction}, pp. 127--136, 2014.

\bibitem{tregillus2017handsfree}
S.~Tregillus, M.~Al~Zayer, and E.~Folmer.
\newblock Handsfree omnidirectional {VR} navigation using head tilt.
\newblock In {\em Proc. of the 2017 CHI Conf. on Human Factors in Computing Systems}, pp. 4063--4068, 2017.

\bibitem{digitaltrends_vr}
D.~Trends.
\newblock Stand up or sit down? many don't take advantage of {VR}'s room-scale experience, 2018.
\newblock Accessed: September 18, 2023.

\bibitem{triantafyllidis2021challenges}
E.~Triantafyllidis and Z.~Li.
\newblock The challenges in modeling human performance in 3{D} space with fitts’ law.
\newblock In {\em Extended Abstracts of the 2021 CHI Conf. on Human Factors in Computing Systems}, CHI EA '21. ACM, USA, 2021. \href{https://doi.org/10.1145/3411763.3443442}
{doi: {{%
10\hspace{.1pt}\discretionary{.}{%
}{.}\hspace{.4pt}1145\discretionary{/}{%
}{/}3411763\hspace{.1pt}\discretionary{.}{%
}{.}\hspace{.4pt}3443442}}}


\bibitem{usoh1999walking}
M.~Usoh, K.~Arthur, M.~C. Whitton, R.~Bastos, A.~Steed, M.~Slater, and F.~P. Brooks.
\newblock Walking > walking-in-place > flying, in virtual environments.
\newblock In {\em Proc. of the 26th Annual Conf. on Computer Graphics and Interactive Techniques (SIGGRAPH '99)}, pp. 359--364. ACM Press/Addison-Wesley Publishing Co., USA, 1999. \href{https://doi.org/10.1145/311535.311589}
{doi: {{%
10\hspace{.1pt}\discretionary{.}{%
}{.}\hspace{.4pt}1145\discretionary{/}{%
}{/}311535\hspace{.1pt}\discretionary{.}{%
}{.}\hspace{.4pt}311589}}}


\bibitem{wagner2023fitts}
U.~Wagner, M.~N. Lystb{\ae}k, P.~Manakhov, J.~E.~S. Gr{\o}nb{\ae}k, K.~Pfeuffer, and H.~Gellersen.
\newblock A fitts’ law study of gaze-hand alignment for selection in 3{D} user interfaces.
\newblock In {\em Proc. of the 2023 CHI Conf. on Human Factors in Computing Systems}, pp. 1--15, 2023.

\bibitem{wang2012comparing}
J.~Wang and R.~W. Lindeman.
\newblock Comparing isometric and elastic surfboard interfaces for leaning-based travel in 3{D} virtual environments.
\newblock In {\em 2012 IEEE Symposium on 3{D} User Interfaces ( 3{D}UI)}, pp. 31--38. IEEE, 2012.

\bibitem{weissker2018spatial}
T.~Wei{\ss}ker, A.~Kunert, B.~Fr{\"o}hlich, and A.~Kulik.
\newblock Spatial updating and simulator sickness during steering and jumping in immersive virtual environments.
\newblock In {\em 2018 IEEE conference on virtual reality and 3{D} user interfaces (VR)}, pp. 97--104. IEEE, 2018.

\bibitem{wilson2016vr}
P.~T. Wilson, W.~Kalescky, A.~MacLaughlin, and B.~Williams.
\newblock Vr locomotion: walking> walking in place> arm swinging.
\newblock In {\em Proc. of the 15th ACM SIGGRAPH Conf. on Virtual-Reality Continuum and Its Applications in Industry-Volume 1}, pp. 243--249, 2016.

\bibitem{wobbrock2011effects}
J.~O. Wobbrock, K.~Shinohara, and A.~Jansen.
\newblock The effects of task dimensionality, endpoint deviation, throughput calculation, and experiment design on pointing measures and models.
\newblock In {\em Proc. of the SIGCHI Conf. on Human Factors in Computing Systems}, CHI '11, p. 1639–1648. ACM, USA, 2011. \href{https://doi.org/10.1145/1978942.1979181}
{doi: {{%
10\hspace{.1pt}\discretionary{.}{%
}{.}\hspace{.4pt}1145\discretionary{/}{%
}{/}1978942\hspace{.1pt}\discretionary{.}{%
}{.}\hspace{.4pt}1979181}}}


\bibitem{wolf2020understanding}
D.~Wolf, J.~Gugenheimer, M.~Combosch, and E.~Rukzio.
\newblock Understanding the heisenberg effect of spatial interaction: A selection induced error for spatially tracked input devices.
\newblock In {\em Proc. of the 2020 CHI Conf. on Human Factors in Computing Systems (CHI '20)}, pp. 1--10. ACM, 2020. \href{https://doi.org/10.1145/3313831.3376876}
{doi: {{%
10\hspace{.1pt}\discretionary{.}{%
}{.}\hspace{.4pt}1145\discretionary{/}{%
}{/}3313831\hspace{.1pt}\discretionary{.}{%
}{.}\hspace{.4pt}3376876}}}


\bibitem{zhai2004speed}
S.~Zhai, J.~Kong, and X.~Ren.
\newblock Speed–accuracy tradeoff in fitts’ law tasks—on the equivalency of actual and nominal pointing precision.
\newblock {\em Int. J. of Human-Computer Studies}, 61(6):823--856, 2004.
\newblock Fitts' law 50 years later: applications and contributions from human-computer interaction. \href{https://doi.org/10.1016/j.ijhcs.2004.09.007}
{doi: {{%
10\hspace{.1pt}\discretionary{.}{%
}{.}\hspace{.4pt}1016\discretionary{/}{%
}{/}j\hspace{.1pt}\discretionary{.}{%
}{.}\hspace{.4pt}ijhcs\hspace{.1pt}\discretionary{.}{%
}{.}\hspace{.4pt}2004\hspace{.1pt}\discretionary{.}{%
}{.}\hspace{.4pt}09\hspace{.1pt}\discretionary{.}{%
}{.}\hspace{.4pt}007}}}


\bibitem{zhang2017double}
F.~Zhang, S.~Chu, R.~Pan, N.~Ji, and L.~Xi.
\newblock Double hand-gesture interaction for walk-through in {VR} environment.
\newblock In {\em 2017 IEEE/ACIS 16th Int. Conf. on Computer and Information Science (ICIS)}, pp. 539--544. IEEE, 2017.

\bibitem{zielasko2016evaluation}
D.~Zielasko, S.~Horn, S.~Freitag, B.~Weyers, and T.~W. Kuhlen.
\newblock Evaluation of hands-free hmd-based navigation techniques for immersive data analysis.
\newblock In {\em 2016 IEEE Symposium on 3{D} User Interfaces ( 3{D}UI)}, pp. 113--119. IEEE, 2016.

\bibitem{zielasko2020take}
D.~Zielasko, Y.~C. Law, and B.~Weyers.
\newblock Take a look around--the impact of decoupling gaze and travel-direction in seated and ground-based virtual reality utilizing torso-directed steering.
\newblock In {\em 2020 IEEE Conference on Virtual Reality and 3{D} User Interfaces (VR)}, pp. 398--406. IEEE, 2020.

\bibitem{zielasko2020can}
D.~Zielasko and B.~E. Riecke.
\newblock Can we give seated users in virtual reality the sensation of standing or even walking? do we want to?
\newblock In {\em 2020 IEEE Conf. on Virtual Reality and 3{D} User Interfaces Abstracts and Workshops (VRW)}, pp. 281--282. IEEE, 2020.

\bibitem{zielasko2020either}
D.~Zielasko and B.~E. Riecke.
\newblock Either give me a reason to stand or an opportunity to sit in {VR}.
\newblock In {\em 2020 IEEE Conf. on Virtual Reality and 3{D} User Interfaces Abstracts and Workshops (VRW)}, pp. 283--284. IEEE, 2020.

\bibitem{zielasko2020sitting}
D.~Zielasko and B.~E. Riecke.
\newblock Sitting vs. standing in {VR}: Towards a systematic classification of challenges and (dis) advantages.
\newblock In {\em VR Workshops}, pp. 297--298, 2020.

\bibitem{zielasko2021sit}
D.~Zielasko and B.~E. Riecke.
\newblock To sit or not to sit in {VR}: Analyzing influences and (dis) advantages of posture and embodied interaction.
\newblock {\em Computers}, 10(6):73, 2021.

\bibitem{zielasko2023stay}
D.~Zielasko and T.~Weissker.
\newblock Stay vigilant: The threat of a replication crisis in {VR} locomotion research.
\newblock In {\em Proceedings of the 29th ACM Symposium on Virtual Reality Software and Technology}, pp. 1--10, 2023.

\end{thebibliography}
